\documentclass[twoside, twocolumn, 9pt]{article}
\usepackage{extsizes}
\usepackage{extsizes}
\usepackage[super,sort&compress,comma]{natbib} 
\usepackage[version=3]{mhchem}
\usepackage[left=1.5cm, right=1.5cm, top=1.785cm, bottom=2.0cm]{geometry}
\usepackage{balance}
\usepackage{times,mathptm}
\usepackage{sectsty}
\usepackage{graphicx} 
\usepackage{lastpage}
\usepackage[format=plain,justification=raggedright,singlelinecheck=false,font={stretch=1.125,small,sf},labelfont=bf,labelsep=space]{caption}
\usepackage{float}
\usepackage{fancyhdr}
\usepackage{fnpos}
\usepackage[english]{babel}
\usepackage{array}
\usepackage{droidsans}
\usepackage{charter}
\usepackage[T1]{fontenc}
\usepackage[usenames,dvipsnames]{xcolor}
\usepackage{setspace}

\usepackage{ulem}
\usepackage{ragged2e}
\usepackage{xr}
\usepackage[compact]{titlesec}
\usepackage{siunitx}

\DeclareCaptionJustification{justified}{\justifying}
\usepackage{latexsym}
\usepackage{amssymb}
\usepackage{amsmath}  
\definecolor{cream}{RGB}{222,217,201}

\usepackage[hyperfootnotes=false]{hyperref}	 
\hypersetup{
  colorlinks,
  citecolor=blue,
  linkcolor=blue,
  urlcolor=blue
}

\newcommand*{\citen}[1]{
  \begingroup
    \romannumeral-`\x 
    \setcitestyle{numbers}%
    \cite{#1}%
  \endgroup   
}
\def\ln{{\operatorname{ln}}}

\def\rmd{{\mathrm{d}}}

\def\rme{{\mathrm{e}}}

\def\Eq{eqn}
\def\Eqs{eqs.}

\def\Fig{Fig.}

\def\SItext{{\em Supporting Information}}

\newcommand{\eqs}{\;\!}

\newcommand{\kB}{k_\mathrm{B}}

\newcommand{\trm}[1]{{\textrm{#1}}}

\usepackage{soul}  

\begin{document}

\pagestyle{fancy}
\thispagestyle{plain}
\fancypagestyle{plain}{

\fancyhead[C]{\includegraphics[width=18.5cm]{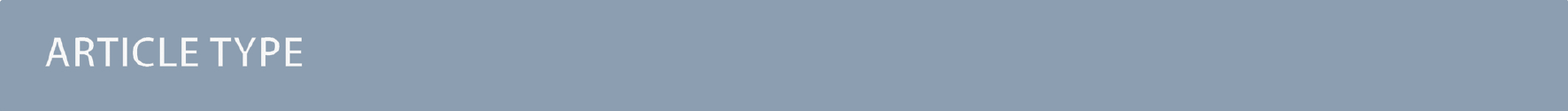}}
\fancyhead[L]{\hspace{0cm}\vspace{1.5cm}\includegraphics[height=30pt]{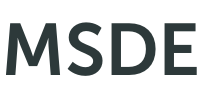}}
\fancyhead[R]{\hspace{0cm}\vspace{1.7cm}\includegraphics[height=55pt]{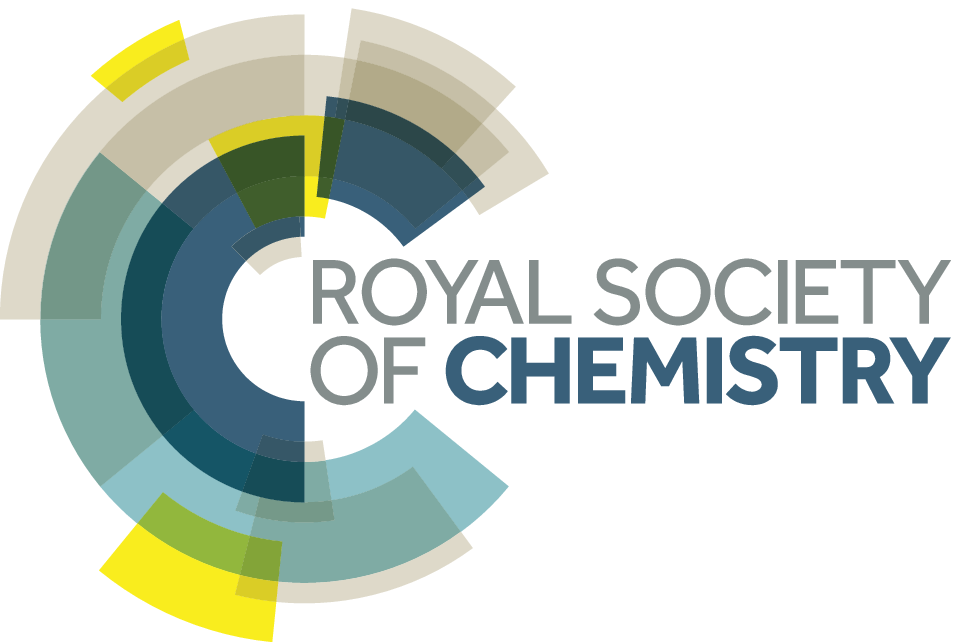}}
\renewcommand{\headrulewidth}{0pt}
}

\makeFNbottom
\makeatletter
\renewcommand\LARGE{\@setfontsize\LARGE{15pt}{17}}
\renewcommand\Large{\@setfontsize\Large{12pt}{14}}
\renewcommand\large{\@setfontsize\large{10pt}{12}}
\renewcommand\footnotesize{\@setfontsize\footnotesize{7pt}{10}}
\makeatother

\renewcommand{\thefootnote}{\fnsymbol{footnote}}
\renewcommand\footnoterule{\vspace*{1pt}%
\color{cream}\hrule width 3.5in height 0.4pt \color{black}\vspace*{5pt}} 
\setcounter{secnumdepth}{5}

\makeatletter 
\renewcommand\@biblabel[1]{#1}            
\renewcommand\@makefntext[1]%
{\noindent\makebox[0pt][r]{\@thefnmark\,}#1}
\makeatother 
\renewcommand{\figurename}{\small{Fig.}~}
\sectionfont{\sffamily\Large}
\subsectionfont{\normalsize}
\subsubsectionfont{\bf}
\setstretch{1.125} 
\setlength{\skip\footins}{0.8cm}
\setlength{\footnotesep}{0.25cm}
\setlength{\jot}{10pt}
\titlespacing*{\section}{0pt}{4pt}{4pt}
\titlespacing*{\subsection}{0pt}{15pt}{1pt}

\fancyfoot{}
\fancyfoot[LO,RE]{\vspace{-7.1pt}\includegraphics[height=9pt]{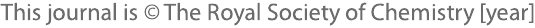}}
\fancyfoot[CO]{\vspace{-7.1pt}\hspace{13.2cm}\includegraphics{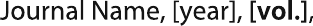}}
\fancyfoot[CE]{\vspace{-7.2pt}\hspace{-14.2cm}\includegraphics{head_foot/RF}}
\fancyfoot[RO]{\footnotesize{\sffamily{1--\pageref{LastPage} ~\textbar  \hspace{2pt}\thepage}}}
\fancyfoot[LE]{\footnotesize{\sffamily{\thepage~\textbar\hspace{3.45cm} 1--\pageref{LastPage}}}}
\fancyhead{}
\renewcommand{\headrulewidth}{0pt} 
\renewcommand{\footrulewidth}{0pt}
\setlength{\arrayrulewidth}{1pt}
\setlength{\columnsep}{6.5mm}
\setlength\bibsep{1pt}

\makeatletter 
\newlength{\figrulesep} 
\setlength{\figrulesep}{0.5\textfloatsep} 

\newcommand{\topfigrule}{\vspace*{-1pt}%
\noindent{\color{cream}\rule[-\figrulesep]{\columnwidth}{1.5pt}} }

\newcommand{\botfigrule}{\vspace*{-2pt}%
\noindent{\color{cream}\rule[\figrulesep]{\columnwidth}{1.5pt}} }

\newcommand{\dblfigrule}{\vspace*{-1pt}%
\noindent{\color{cream}\rule[-\figrulesep]{\textwidth}{1.5pt}} }

\makeatother

\twocolumn[
  \begin{@twocolumnfalse}
  
\vspace{3cm}
\sffamily
\begin{tabular}{m{4.5cm} p{13.5cm} }

\includegraphics{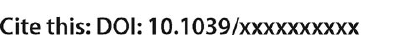} & \noindent\LARGE{\textbf{Modeling of stimuli-responsive nanoreactors: rational rate control towards the design of colloidal enzymes}} \\
\vspace{0.3cm} & \vspace{0.3cm} \\

& \noindent\large{Matej Kandu\v{c},$^\trm{a}$ Won Kyu Kim,$^\trm{b}$ Rafael Roa$^\trm{c}$ and Joachim Dzubiella$^\trm{de}$}\\ 


\includegraphics{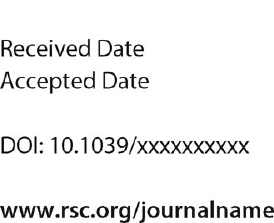} & \noindent\normalsize{

In modern applications of heterogeneous liquid-phase nanocatalysis, the catalysts (e.g., metal nanoparticles) need to be typically affixed to a colloidal carrier system for stability and easy handling. ``Passive carriers'' (e.g., simple polyelectrolytes) serve for a controlled synthesis of the nanoparticles and prevent coagulation during catalysis. Recently, however, hybrid conjugates of nanoparticles and synthetic thermosensitive polymers have been developed that enable to change the catalytic activity of the nanoparticles by external triggers. In particular, nanoparticles embedded in a stimuli-responsive network made from poly(N-isopropylacrylamide) (PNIPAM) have become the most-studied examples of such hybrids. It has been demonstrated that the permeability of the polymer network and thus the reactant flux can be switched and controlled by external stimuli.  Such ``active carriers'' may thus be viewed as true nanoreactors that open up new design routes in nano-catalysis and elevate synthesis to create highly selective, programmable ``colloidal enzymes''.  However, only a comprehensive understanding of these materials on all time and length scales can lead to a rational design of future, highly functional materials. Here we review the current state of the theoretical and multi-scale simulation approaches, aiming at a fundamental understanding of these nanoreactors. In particular, we summarize a theoretical  approach for reaction rates of surface-catalyzed bimolecular reactions in responsive nanoreactors in terms of the key material parameters, the polymer shell permeability $\mathcal{P}$ and the reactant partition ratio $\mathcal K$. We discuss recent computer simulation studies of both atomistic and coarse-grained polymer models in which these quantities have been characterized in some detail. We conclude with an outlook on selected open questions and future theoretical challenges in nanoreactor modeling.
 
} \\

\end{tabular}

 \end{@twocolumnfalse} \vspace{0.6cm}

  ]

\renewcommand*\rmdefault{bch}\normalfont\upshape
\rmfamily
\section*{}
\vspace{-1cm}



\footnotetext{\textit{$^\trm{a}$~Jo\v{z}ef Stefan Institute, Jamova 39, SI-1000 Ljubljana, Slovenia}}

\footnotetext{\textit{$^\trm{b}$~Korea Institute for Advanced Study, 85 Hoegiro, Seoul 02455, Republic of Korea.}}

\footnotetext{\textit{$^\trm{c}$~Departamento de F\'{i}sica Aplicada I, Facultad de Ciencias, Universidad de M\'{a}laga, Campus de Teatinos s/n, E-29071 M\'{a}laga, Spain}}

\footnotetext{\textit{$^\trm{d}$~Research Group for Simulations of Energy Materials, Helmholtz-Zentrum Berlin, Hahn-Meitner-Platz~1, D-14109 Berlin, Germany}}

\footnotetext{\textit{$^\trm{e}$~Institut f\"ur Physik, Albert-Ludwigs-Universit\"{a}t Freiburg, Hermann-Herder-Str.~3, D-79104 Freiburg, Germany}}





\section{Introduction}

Synthetic nanoreactors are an emerging and promising new nanotechnology for liquid-phase heterogenous catalysis.  In these nanoreactors, the catalysts are confined in a permeable nanostructure, which acts as a carrier and can be used to shelter and control the catalytic processes. In particular,  the catalysis can be made selective and responsive if the nanoreactor permeability differentiates among molecules and can be modulated by external stimuli.~\cite{Petrosko:2016he,Stuart:2010hu,Campisi:2016hl,Lu:2011bi,CarregalRomero:2010gp,Herves:2012fp,Wu:2012bx,Jia:2016cy,Prieto:2016bj,Gaitzsch:2015kr,Vriezema:2005fx,Renggli:2011if,Tanner:2011jf,Guan:2011eva}
These nanoreactors can be used for a large variety of applications, ranging from analytical tools to study chemical reactions~\cite{Petrosko:2016he,Lu:2011bi,CarregalRomero:2010gp,Herves:2012fp,Wu:2012bx,Jia:2016cy,Prieto:2016bj,Vriezema:2005fx,Renggli:2011if,Campisi:2016hl,Stuart:2010hu,Gaitzsch:2015kr} to biosensors for the diagnosis of diseases.~\cite{Vriezema:2005fx,Renggli:2011if,Tanner:2011jf,Gaitzsch:2015kr,Guan:2011eva}
Examples of natural nanoreactors are lipid-based membranes ({e.g.}, liposomes), cage-like proteins ({e.g.}, ferritins), protein-based bacterial microcompartments, and viruses.~\cite{Vriezema:2005fx,Renggli:2011if,Tanner:2011jf,Liu:2016gf}
Artificial nanoreactors (based on spherical polyelectrolyte brushes, dendrimers, ligands, or even DNA) are simpler than the natural ones and thus easier to control for targeted applications.~\cite{Lu:2011bi,CarregalRomero:2010gp,Herves:2012fp,Wu:2012bx,Jia:2016cy,Prieto:2016bj,Vriezema:2005fx,Renggli:2011if,Montolio:2016jy,Zinchenko:2016jy,Gaitzsch:2015kr} 

In particular, nanoreactors containing metal nanoparticles have emerged as a promising catalytic system.~\cite{Lu:2011bi,Herves:2012fp,Wu:2012bx,Prieto:2016bj,Lu:2006cr,Zhang:2010hx,ContrerasCaceres:2008hk,CarregalRomero:2010gp,Jia:2016cy,Li:2016bs} For example, gold becomes an active catalyst when divided down to the nanophase.~\cite{Haruta2003,Hutchings2005,Zhang2012,Astruc2008ed,Zhao2013,Li2014ange}
However, the handling of the particles in the liquid phase is an important problem: The surface of the particles should be easily accessible for the mixture of the reactants. This condition would require the nanoparticles to be freely suspended in the solution, and coagulation or any type of Ostwald ripening of the nanoparticles should not occur during the catalytic reaction. Also, leaching of metal or loss of nanoparticles from the carrier must be prevented to ensure a meaningful and repeated use of the catalyst. The latter requirements necessitate a suitable carrier that ensures a safe and repeated handling of the nanoparticles.~\cite{Boisselier2009,Taylor2014} It was demonstrated that suitable carrier systems include colloidal particles,\cite{Sharma2004,Mei2005} dendrimers,~\cite{Antonels2013,Crooks2001a,Anderson2015,Deraedt2014, ESUMI2002402,Bingwa2014,BINGWA20151,NOH2014400,NOH2015107, Pozun2013,Johnson2013} mesoporous materials,~\cite{Gross2014,Calvo2010,Brock1998} spherical polyelectrolyte brushes,~\cite{Ballauff2006cur,Ballauff2007prog} and other systems~\cite{Cao2015} structured on a length scale between one and a few hundred nm. 

\begin{figure}[t!]
\begin{center}
\includegraphics[width=0.65\linewidth]{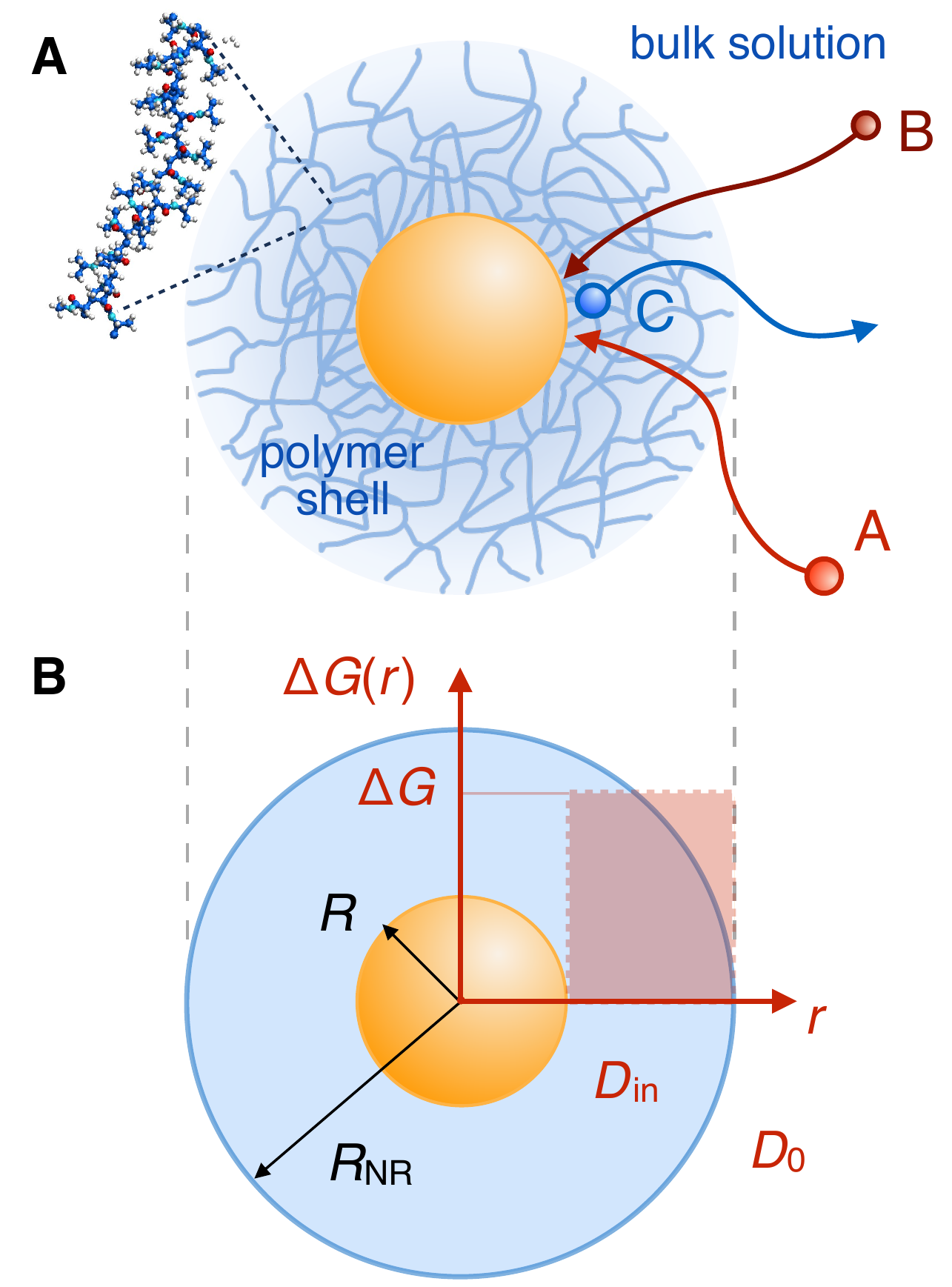}
\caption{Bimolecular reactions in core-shell nanoreactors. 
(A) Two reactants, A and B, diffusing from a bulk solution, generate a product, C, in the proximity of a catalyst nanoparticle (central yellow sphere) embedded in a PNIPAM polymer network.
(B) Schematic representation of a core-shell nanoreactor. A nanoparticle of radius $R$ is embedded in a spherical shell of outer radius $R_\mathrm{NR}$. The shell permeability depends on the diffusivity, $D(r)$, and on the transfer free energy profiles, $\Delta G(r)$. We model both as step functions with values $D_\mathrm{in}$ and $\Delta G$ inside, and $D_0$ and zero reference outside the shell, respectively.} 
\label{fig_sketch}
\end{center}
\end{figure}

In recent years, the concept of such carrier systems has been further advanced with the synthesis of hydrogel-based nanoreactors, for which rate control by external stimuli has become possible.~\cite{Lu:2006cr,Herves:2012fp,Wu:2012bx,Renggli:2011if,Gaitzsch:2015kr,Lu2008ma,BALLAUFF20071815,Lu2009jma,Wu2012indu,LU2013639,resmini2010microgels,Jia2015jmat,Liu:2016gf,Welsch2009,Lu2006ange}  Thermosensitive hydrogels made from a network of poly(N-isopropylacrylamide) (PNIPAM) and its copolymers provide a good example:~\cite{Lu:2006cr,Zhang:2010hx,Lu:2011bi,Herves:2012fp,Wu:2012bx,ContrerasCaceres:2008hk,CarregalRomero:2010gp,Jia:2016cy,Li:2016bs,Lu2006jpcb,yang2015amphiphilic,Shi2014,Chang2015ch,LIU20151,tang2015synthesis,Plazas2015,Wu2015chem,Plamper2017,Gu:2014cf} 
Typical colloidal carrier architectures are of core-shell or yolk-shell type where the polymer gel constitutes a permeable shell around a solid core (core-shell) or around a hollow void (yolk-shell).\cite{Herves:2012fp} The core can be the nanoparticle itself, cf.~Fig.~\ref{fig_sketch}, or, for example, a polystyrene core. The catalytic nanoparticles can be located during synthesis in a well controlled fashion, e.g.,  into the voids, onto the cores, or distributed within the polymer shell. 
Sometimes simply a pure hydrogel (nano- or microgel) particle is the carrier for the catalysts.  In this case carrier and polymer shell in our context are essentially the same. A survey of selected but very typical experimental architectures and results for polymer-based nanoreactor carriers is provided in Tab.~\ref{tab:table1}.

\begin{table*}[t!]
\small
  \begin{center}
    \caption{Survey of selected publications on responsive nanoreactor catalytic experiments with different architectures. 
    Polymer abbreviations: poly(N-isopropylacrylamide) (PNIPAM), polysterene (PS), maleated carboxymethylchitosan (MACACS), poly(N-vinylcaprolactam) (PVCL), poly(styrene-NIPAM) (P(S-NIPAM)), poly(NIPAM-co-methacrylic acid) (P(NIPAM-co-MAA)), poly(NIPAM-co-2-(dimethylamino)ethyl methacrylate) (P(NIPAM-co-AMPS)), poly(4-vinylphenylboronic acid-co-NIPAM-co-acrylamide) (P(VPBA-NIPAM-AAm)).
Solute abbreviations: 4-nitrophenol (NP), 4-aminophenol (AP), nitrobenzene (NB), aminobenzene (AB), hexacyanoferrate(III) (HCF), hexacyanoferrate(II) (HCF2), o-nitrophenyl-D-glucopyranoside (oNPG).}
    \label{tab:table1}
\resizebox{\textwidth}{!}{
    \begin{tabular}{|p{0.4cm}|p{1.4cm}|p{2.9cm}|p{1.2cm}|p{5.8cm}|p{8cm}|}
      \hline
      Ref. & Architecture & Core - Polymer & Catalyst & Reaction & Result \\
      \hline      
      \citen{Wu:2012bx}	 & 	Yolk-shell	 & 	Au - PNIPAM	 & 	Au	 & 	reduction: NP $\rightarrow$ AP \ \& \ NB $\rightarrow$ AB	 & 	$T$-dependence on gel swelling and reaction rate	 \\ 
      \citen{CarregalRomero:2010gp},\citen{Herves:2012fp}	 & 	Core-shell	 & 	Au - PNIPAM	 & 	Au	 & 	reduction: HCF $\rightarrow$ HCF2	 & 	$T$-dependence on gel swelling and reaction rate, rate dependence on nanoreactor concentration and cross-linking density	 \\ 
      \citen{Herves:2012fp},\citen{Gu:2014cf}	 & 	Core-shell	 & 	Pt/Au - PNIPAM	 & 	Pt/Au	 & 	reduction: NP $\rightarrow$ AP	 & 	rate dependence on reactant concentration	 \\ 
      \citen{Jia:2016cy}	 & 	Core-shell	 & 	Cu$_2$O - PNIPAM	 & 	Cu$_2$O	 & 	decomposition by visible light: methyl orange	 & 	$T$-dependence on gel swelling and reaction rate	 \\ 
      \citen{Li:2016bs}	 & 	Core-shell	 & 	Au - PNIPAM	 & 	Ag	 & 	reduction: NP $\rightarrow$ AP	 & 	photoresponsive gel size and reaction rate	 \\ 
      \citen{Wu2015chem}	 & 	Core-shell	 & 	Au - P(VPBA-NIPAM-AAm)	 & 	Au	 & 	reduction: NP $\rightarrow$ AP \ \& \ NB $\rightarrow$ AB	 & 	glucose concentration dependence on gel swelling and reaction rate	 \\ 
      \citen{Lu:2011bi},\citen{Lu2009jma}	 & 	Core-shell	 & 	PS - PNIPAM	 & 	Au/Pt/Rh	 & 	oxidation: benzyl alcohol $\rightarrow$ benzaldehyde	 & 	$T$-dependence on gel swelling and reaction rate	 \\ 
      \citen{Lu:2011bi},\citen{Welsch2009}	 & 	Core-shell	 & 	PS - PNIPAM	 & 	$\beta$-D-glucosidase	 & 	hydrolysis: oNPG $\rightarrow$ D-glucose + o-nitrophenol	 & 	$T$-dependence on gel swelling and reaction rate	 \\ 
      \citen{Lu2006ange}	 & 	Core-shell	 & 	PS - PNIPAM	 & 	Ag	 & 	reduction: NP $\rightarrow$ AP	 & 	$T$-dependence on gel swelling and reaction rate	 \\ 
      \citen{yang2015amphiphilic}	 & 	Core-shell	 & 	P(S-NIPAM) - P(NIPAM-co-MAA)	 & 	Ag	 & 	reduction: NP $\rightarrow$ AP \ \& \ NB $\rightarrow$ AB	 & 	$T$-dependence on gel swelling and reaction rate	 \\ 
      \citen{Zhang:2010hx}	 & 	Microgel	 & 	PNIPAM/MACACS	 & 	Ag	 & 	reduction: NP $\rightarrow$ AP	 & 	$T$-dependence on gel swelling and reaction rate	 \\ 
      \citen{Jia2015jmat}	 & 	Microgel	 & 	PVCL-$\alpha$-cyclodextrin	 & 	Au	 & 	reduction: NP $\rightarrow$ AP	 & 	$T$-dependence on gel swelling and reaction rate	 \\ 
      \citen{Shi2014}	 & 	Hydrogel	 & 	P(NIPAM-co-MAA)	 & 	Au	 & 	reduction: NP $\rightarrow$ AP	 & 	$T$-dependence on gel swelling and reaction rate	 \\ 
      \citen{Chang2015ch}	 & 	Microgel	 & 	cellulose	 & 	cellulase (enzyme)	 & 	hydrolysis: cellulose $\rightarrow$ glucose	 & 	$T$-dependence on gel swelling and time-dependent product concentration	 \\ 
      \citen{LIU20151}	 & 	Microgel	 & 	P(NIPAM-co-AMPS)	 & 	Ni	 & 	reduction: NP $\rightarrow$ AP	 & 	$T$-dependence on gel swelling and reaction rate	 \\ 
      \citen{tang2015synthesis}	 & 	Microgel	 & 	P(NIPAM-co-AMPS)	 & 	Ag	 & 	reduction: methylene blue	 & 	$T$-dependence on gel swelling and reaction rate, pH-dependence on gel swelling	 \\      
            \hline
    \end{tabular}
    }
  \end{center}
\end{table*}

The responsive polymer shell is in a swollen hydrophilic state at low temperature, but sharply collapses into a rather hydrophobic state above the critical solution temperature.~\cite{Pelton:2000vo} The sharp volume transition of the gel is reversible~\cite{khokhlov1980swelling,erman1986critical,khokhlov1993conformational,barenbrug1995highly,heskins1968solution,duvsek1968transition, Habicht2015} and depends on the temperature,~\cite{zhou1996situ,wu1997volume} or more general, solvent quality. This has substantial influence on reactant partitioning close to the catalysts as well as reactant transport towards it.~\cite{stefano2015} Hence, there are two key roles of the polymer shell. On the one hand, the shell acts as a integral part of the whole  carrier that protects nanoparticles from aggregation and hinders chemical degradation processes, e.g., oxidation.~\cite{Jia:2016cy} On the other hand, the polymer ability to switch between states with different physicochemical properties upon changes in environmental parameters, e.g., temperature, pH, or concentration of certain solutes, provides a handle to actively control the nanoreactor's catalytic properties.~\cite{stefano2015}

A quantitative study and understanding of a nanoreactor requires kinetic data measured with the highest precision possible. Up to now, most of the testing of the catalytic activity of nanoparticles in aqueous phase has been done using the reduction of 4-nitrophenol by borohydride. Pal et al.~\cite{PRADHAN2002247} and Esumi et al.~\cite{ESUMI2002402} have been the first who have demonstrated the usefulness of this reaction. In the meantime, the reduction of 4-nitrophenol has become the most used model reaction for the quantitative testing and analyzing of the catalytic activity of nanoparticles in the liquid phase.~\cite{Aditya2015,Zhao2015coord}
Further examples of catalytic reactions in aqueous solution studied in this system are the reductions of nitrobenzene and hexacyanoferrate (III) by borohydride ions~\cite{Lu:2006cr,Herves:2012fp,Wu:2012bx,Zhang:2010hx} and the decomposition of methyl orange under visible light.~\cite{Jia:2016cy}

All the aforementioned examples deal with surface-catalyzed {\it bimolecular} reactions, being a very common type. As pointed out before,~\cite{CarregalRomero:2010gp,Herves:2012fp,Wu:2012bx, stefano2015} pseudo-unimolecular surface-catalyzed reactions in responsive nanoreactors can be described by combining a thermodynamic two-state model for the polymer volume transition with the appropriate reaction--diffusion equations.  In particular, the important effect of a change in the shell permeability on the reactants approach to the catalyst's surface can be described by theory of diffusion through an energy landscape,\cite{Wu:2012bx, stefano2015, Galanti:2016bz, Rafa2017} in the spirit of Debye--Smoluchowski diffusion-controlled rate theory.~\cite{Smoluchowski:1917wh,Debye:1942cv,Wilemski:1973ki,Calef:1983vd,Hanggi:1990en} Importantly, the latter also takes into account the local reactant concentration, i.e., the partitioning inside the polymer shell close to the catalyst.  Recently, we have presented an extended theory of diffusion-limited bimolecular reactions in nanoreactors, which can be employed to rationally design the activity and selectivity of a nanoreactor.~\cite{stefano2015,Rafa2017}
The main result of our consideration was the following formula for the total catalytic rate in bimolecular reactions in core-shell nanoreactors (cf.~Fig.~\ref{fig_sketch}):
\begin{align}\label{equa1}
k_\mathrm{tot}=
&\frac{1}{2}
\Bigg[
\frac{k_{D^\mathrm{A}}k_{D^\mathrm{B}}}{k_\mathrm{R}}+k_{D^\mathrm{A}}+k_{D^\mathrm{B}}
\nonumber \\
&-\sqrt{\left(
\frac{k_{D^\mathrm{A}}k_{D^\mathrm{B}}}{k_\mathrm{R}}+k_{D^\mathrm{A}}+k_{D^\mathrm{B}}
\right)^2-4k_{D^\mathrm{A}}k_{D^\mathrm{B}}}
\Bigg].
\eqs
\end{align}
Here, $k_{D^\mathrm{A}}(\mathcal{P}^\mathrm{A})$ and $k_{D^\mathrm{B}}(\mathcal{P}^\mathrm{B})$ are the diffusion rates of the reactants A and B, which {\it explicitly} depend on the shell permeability $\mathcal{P}^i$, and $k_\mathrm{R}(\mathcal{K}^\mathrm{A},\mathcal{K}^\mathrm{B})$ is the surface reaction rate, {\it explicitly} depending on partitioning $\mathcal{K}^i$ as defined below.

In general, permeability of a material defines the ability of the penetrating molecules (e.g., gas, ligands, reactants, etc.) to permeate and flow through a given medium under the action of an external field or chemical gradient.  It is thus without doubt one of the most fundamental transport descriptors employed in the physical sciences and material engineering.  In the standard `solution--diffusion' picture for permeable membranes,  it is commonly defined on the linear response level by~\cite{yasuda1969permeability3,robeson, Baker, diamond1974, palasis1992permeability, gehrke, chauhan,Thomas2001transport,Ulbricht2006, Baker2014,park2017} 
\begin{equation}\label{eq:P}
\mathcal{P} = \mathcal{K} D_{\rm in},
\end{equation}
where 
\begin{equation}
\mathcal{K} = \frac{c_{\rm in}}{c_{\rm 0}}
\label{eq:Kdirect}
\end{equation}
 is the partition ratio, in this work simply referred to as {\it partitioning}, defined as the ratio of number densities of the solutes inside ($c_{\rm in}$) and outside ($c_{\rm 0}$) the medium in equilibrium, and $D_\text{in}$ is the diffusion coefficient of those inside. Permeability can be thus defined as the inverse of a diffusional resistance of a medium regarding the total mass transport (flux) towards the catalyst driven by the reaction. 
The optimization of permeability, especially for being highly selective among different solutes, has been a grand challenge in material design over the last decades.~\cite{prevost:2007, chauhan, park2017} Prominent applications revolve around gas separation and recovery,~\cite{robeson, chauhan, MOF, lyd, obliger, park2017} desalination and nanofiltration (`molecular sieving'),~\cite{shannon,geise2011,tansel} medical treatments by dialysis or selective drug transport,~\cite{peppas1999, stamatialis} hydrogel-based soft sensors, and the nanoreactors.~\cite{palasis1992permeability, Stuart:2010hu, Lu:2011bi, Rafa2017} We have studied partitioning and permeability of polymer networks and PNIPAM polymers recently on the molecular level by coarse-grained~\cite{kim2017cosolute,kim2019tuning} as well as all-atom molecular dynamics computer simulations.~\cite{kanduc2017selective, milster2019crosslinker, kanduc2018diffusion, kanduc2019free}

Hence, in the last couple of years many quantitative concepts have emerged both on the continuum and the microscopic level that will eventually lead to a more fundamental understanding of nanoreactors in the future. The possibility arises to optimize responsive nanoreactors to reach the high recognition, selectivity, and feedback control as found for enzymes,~\cite{enzyme,enzyme2} to create ``colloidal enzymes''. Here we review the state-of-the-art of the current understanding of the intricate links between nanoreactor reaction rate and polymer permeability.  Most of the results presented here are based on our recent research endeavor of multi-scale modeling schemes of hydrogel systems in order to establish rational design principles of responsive nanoreactors. We start in Section 2 by summarizing the rate theory for nanoparticle-catalyzed bimolecular reactions including partitioning and permeability of the polymer shell. The key property to be tuned and `programmed' during the synthesis in order to select and switch catalytic activity is the permeability.  In Section 3 we thus proceed with mesoscale coarse-grained computer models, which give fundamental insights on partition--diffusion correlations in the permeability and how they can be tuned qualitatively by microscopic interactions. In the last part we turn to atomistically-resolved molecular simulations of the PNIPAM hydrogel models in swollen and collapsed state.  Here we address the question of the influence of the `chemistry' of the interactions, e.g., role of (temperature-dependent) hydration, polarity, reactant type and size, etc. Obviously, there are many open questions, missing connections, and remaining challenges to overcome to obtain a comprehensive multi-scale model. We will briefly discuss those and give an outlook in the final, concluding section. 

\section{Bimolecular reactions in nanoreactors}

\subsection{Macroscopic rates and dependence on permeability}

We review the rate theory for nanoreactors for the case of surface-catalyzed bimolecular reactions in one of the simplest nanoreactor geometries, a core-shell nanoreactor,\cite{Rafa2017} depicted in Fig.~\ref{fig_sketch}A, where a catalytically active metal nanoparticle of radius $R$ is embedded in a thermoresponsive hydrogel matrix of outer radius $R_\mathrm{NR}$. In this spherically symmetric system, we consider that two species A and B diffuse from a bulk solution with respective (initial) concentration $c^\mathrm{A}_0$ and $c^\mathrm{B}_0$ through the polymer shell towards the catalyst nanoparticle. A fraction of the reactants arriving at the surface combines with each other to produce a third molecular species C. Assuming a total concentration of nanoreactors $c_{\rm NR}$, the experimentalist would measure the transformation of a reactant (say reactant A) per time, according to  
\begin{eqnarray}
\frac{{\rm d}c^\trm{A}(t)}{{\rm d}t} = - k_{\rm tot}(c^\trm{A}(t), c^\trm{B}(t)) c_{\rm NR}
\label{exp}
\end{eqnarray}
with instantaneous bulk concentrations $c^\trm{A}(t)$ and $c^\trm{B}(t)$, and $k_{\rm tot}$ has the units of inverse time and is a non-trivial function of the reactant concentrations. In general, and as we will see below, the chemical reaction has no well-defined order. In some limits, e.g., in an abundance of species B, it may reduce to pseudo unimolecular or even pseudo first-order kinetics~\cite{Rafa2017} (see also Section 2.2 later). 

To derive the functional form of the total catalytic rate, we assume $k_\mathrm{tot}$ (number of molecules reacting per unit of time) is equal to the radial flux of reactants at the nanoparticle surface. In bimolecular reactions, then the fraction of molecules A reacting is proportional to the number of molecules B at the same location, and vice versa. Thus, $k_\mathrm{tot}$ can be estimated through the standard mean-field relation~\cite{stefano2015,Atkins} 
\begin{equation}\label{eq:ktotdef}
k_\mathrm{tot}= K_\mathrm{vol} c^\mathrm{A}(R) c^\mathrm{B}(R)
\eqs,
\end{equation}
where $c^\mathrm{A}(R)$ and $c^\mathrm{B}(R)$ are the reactant concentrations at the nanoparticle surface, and $ K_\mathrm{vol}$ the probability that the two species react on the surface (with units per time and per concentration squared).
To calculate $k_\mathrm{tot}$, we solve the stationary continuity equation for the density fields of reactants,
\begin{equation}\label{eq:cont}
\nabla \cdot \mathbf{J}^i =0
\eqs,
\end{equation}
with $\mathbf{J}^i(r)$ being the radial flux of the species $i=$~A, B, C as a function of the distance from the nanoparticle. We make the stationarity assumption that the system is always in a steady-state and there is no explicit time-dependence of the fluxes. In other words, we assume the microscopic relaxation of the system, roughly given by the time of reactants to diffuse through the nanoreactor $R^2_{\rm NR}/D_0$, is faster than the reaction time as defined in \Eq~(\ref{exp}). If we use the fastest, diffusion-controlled (Smoluchowski) rate 
$k_\mathrm{tot} \simeq k_D^0 = 4\pi R D_0  c_0$ as the reaction rate scale, we find the condition for stationarity that 
$c_0 \ll 1/(4\pi R_{\rm NR}^2 R) \simeq 10^{-5}$~mol/l  for typical geometries where $R\simeq 1$~nm and $R_{\rm NR} \simeq 10^2$~nm. In experiments, typically sub-micromolar reactant concentrations are used and the reaction rate is at least 1--2 orders slower than the fastest, fully diffusion-controlled limit, so that the condition is in most cases very well satisfied.

In their diffusive approach to the catalyst nanoparticle, reactants have to permeate the shell. The kinetics of this process is thus governed by the shell permeability, which depends on the diffusivity profile, $D^i(r)$, and on the thermodynamic barrier, i.e., the transfer free energy between bulk and shell, $\Delta G^i(r)$.
For simplicity, we take both profiles to be shell-centered step functions of the width equal to the polymer shell width (see Fig.~\ref{fig_sketch}B), {i.e.}
\begin{equation}\label{eq:difuprofile}
D^i(r)=
\begin{cases}
D_\mathrm{in}^i 	& \text{ \ } R \leq r \leq R_\mathrm{NR}  \eqs,\\
D_0^i 	& \text{ \ } \mathrm{elsewhere} \eqs,
\end{cases}
\end{equation}
and
\begin{equation}\label{eq:DGprofile}
\Delta G^i(r)=
\begin{cases}
\Delta G^i 	& \text{ \ } R \leq r \leq R_\mathrm{NR}  \eqs,\\
0 	& \text{ \ } \mathrm{elsewhere}\eqs.
\end{cases}
\end{equation}
Here, $D_\mathrm{in}^i $ and $D_0^i$ stand for the diffusion coefficients in the polymer shell and solution, respectively. $\Delta G^i$ represents the transfer free energy from bulk water into the shell and as such strongly depends on the state (swollen/collapsed) of the nanoreactor. 
Using standard thermodynamic relations, we connect the flux of the species $i$ to its local concentration $c^i(r)$
\begin{equation}\label{eq:fluxes}
\mathbf{J}^i=-D^i \;\! c^i\nabla\beta\mu^i 
\eqs,
\end{equation}
where $\mu^i(r)$ is the chemical potential of the species $i$, and $\beta=1/\kB T$, with $\kB$ denoting the Boltzmann's constant and $T$ the absolute temperature of the system. 
The chemical potential of a molecule interacting with an external environment with a spatially dependent concentration and free energy is
\begin{equation}\label{eq:chempot}
\beta\mu^i=\ln\left(\frac{c^i}{c_\mathrm{ref}^i}\right)+\beta\Delta G^i
\eqs,
\end{equation}
where $c_\mathrm{ref}^i$ is a reference concentration whose value can be chosen arbitrarily.  Equation~(\ref{eq:chempot}) can now be used to relate the transfer free energy $\Delta G^i$ and partitioning, \Eq~(\ref{eq:Kdirect}),
\begin{equation}\label{eq:partitioning}
\mathcal{K}^i=\exp\left(-\beta\Delta G^i\right)
\eqs.
\end{equation}
With the aforementioned definitions, the shell permeability to the species $i$ is calculated as
\begin{eqnarray}
\mathcal{P}^i= \mathcal{K}^i D_\mathrm{in}^i
\end{eqnarray}
where for $r>R_{NR}$ we have $\mathcal{P}^i = D_0^i$. 

We found~\cite{Rafa2017} that the total catalytic rate for bimolecular reactions in responsive nanoreactors is obtained as in \Eq~(\ref{equa1}).
In this expression,
\begin{equation}\label{eq:surfacerate}
k_\mathrm{R}=K_\mathrm{vol} 
c_0^\mathrm{A} \rme^{-\beta \Delta G^\mathrm{A}(R)}
c_0^\mathrm{B} \rme^{-\beta \Delta G^\mathrm{B}(R)} = K_\mathrm{vol} c_0^\mathrm{A} c_0^\mathrm{B} \mathcal{K}^A \mathcal{K}^B 
\end{equation}
stands for the surface-part of the reaction rate, which is explicitly partitioning-dependent, and
\begin{equation}\label{eq:difurateperm}
k_{D^i}=
4\pi c_0^i
\left[
\int^{\infty}_{R}
\frac{1}{\mathcal{P}^i(r) \;\! r^2}\rmd r
\right]^{-1}
\end{equation}
is the permeability-dependent diffusion part of the reaction rate of the reactant $i$. In the absence of the shell, $\mathcal{P}^i(r) = D_0^i$, and the diffusion rate turns into the Smoluchowski rate $k_{D^i}^0=4\pi R D_0^i c_0^i$.
For the core-shell configuration depicted in Fig.~\ref{fig_sketch}B the step profiles in \Eq~(\ref{eq:difuprofile}) and \Eq~(\ref{eq:DGprofile}) apply and the relation between the shell permeability and the diffusion rate, \Eq~(\ref{eq:difurateperm}), simplifies to 
\begin{equation}\label{eq:kDkD0}
\frac{k_{D^i}}{k_{D^i}^0}
=
\left[
1+\left(\frac{D_0^i}{\mathcal{P}^i}-1\right)
\left(1-\frac{R}{R_\mathrm{NR}}\right) 
\right]^{-1}
\!.
\end{equation}

\begin{figure}[t!]
\begin{center}
\includegraphics[width=0.95\linewidth]{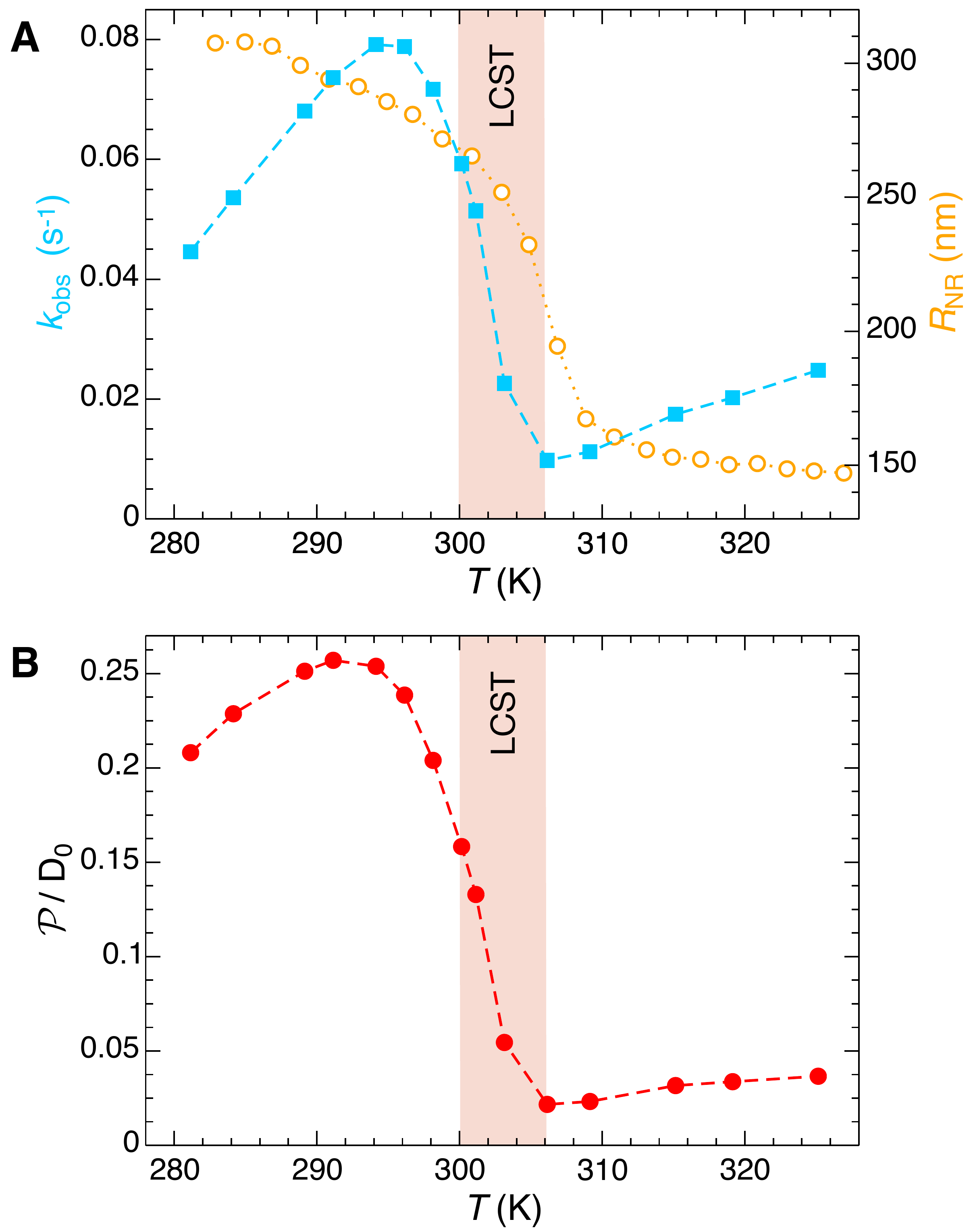}
\caption{
(A)~Temperature dependence of the measured pseudo-first-order constant $k_\trm{obs}$  (blue filled squares) of the electron-transfer reaction between HCF and borohydride ions in Au-PNIPAM nanoreactors.~The measured temperature dependence of the nanoreactor hydrodynamic radius is shown by orange opened circles.~All data were taken from Ref.~\citenum{CarregalRomero:2010gp}. 
(B)~Temperature dependent nanoreactor shell permeability for HCF estimated using \Eq~(\ref{eq:kDkD0}). 
}
\label{figrateperm}
\end{center}
\end{figure}

Equation~(\ref{equa1}) is the main analytical result for nanoparticle surface-catalyzed bimolecular reactions. It shows that, in the fully bimolecular case, the diffusional fluxes of the different reactants are {\it coupled}. Thus, $k_\mathrm{tot}$ depends in a non-trivial way on the surface and the diffusion rates and nanoreactor shell permeability, in contrast to the simple unimolecular case ({i.e.}, in general $k_\mathrm{tot}^{-1}\neq k_D^{-1}+k_\mathrm{R}^{-1}$ in bimolecular reactions).

Equation~(\ref{equa1}) together with~\Eqs~(\ref{eq:surfacerate}) and~(\ref{eq:kDkD0}) can be used to predict the total catalytic rate once the nanoreactor shell permeability and the reactant partition ratios are known (e.g., from experiments, or from simulations, see Sections 3 and 4), or, conversely, to extract the parameters by fitting to experimental data. 
Carregal-Romero {et al.}~\cite{CarregalRomero:2010gp} investigated the bimolecular electron-transfer reaction between hexacyanoferrate (III), $\mathrm{Fe(CN)}_6^{3-}$ (HCF), and borohydride $\mathrm{BH}_4^-$ ions in Au-PNIPAM core-shell nanoreactors.
In a previous work~\cite{Rafa2017} we demonstrated that that bimolecular reaction is diffusion-controlled and can be treated as pseudo-unimolecular (see also next Section 2.2), that is, ${{\rm d}c^\trm{HCF}(t)}/{{\rm d}t} = - k_{\rm obs} c^\trm{HCF}(t)$. 

The temperature dependence of the measured pseudo-first-order constant is shown by blue filled squares in Fig.~\ref{figrateperm}A. We observe that the reaction rate decreases by one order of magnitude when the temperature of the solution exceeds the lower critical solution temperature (LCST) of the PNIPAM polymer. 
The measured nanoreactor hydrodynamic radius data, displayed by orange open circles in Fig.~\ref{figrateperm}A, exhibit the well-known volume transition between the swollen and the collapsed states below and above the LCST, respectively. As we pointed out before, this transition changes the physicochemical properties of the polymer, which leads to different reactant diffusivity and transfer free energy values resulting in a nanoreactor permeability switch at the LCST. 
In the diffusion-controlled limit for pseudo-unimolecular kinetics we can identify $k_{\rm obs} = k_D^{\rm HCF}c_{\rm NR}/c_0^{\rm HCF}$. The $c_0^{\rm HCF}$ is the initial bulk concentration, which can be replaced by the instantaneous $c^{\rm HCF}(t)$ in the equations during the reaction because of the stationarity assumption. Using  \Eq~(\ref{eq:kDkD0}) together with the experimental data from Fig.~\ref{figrateperm}A, we estimate the temperature dependence of the nanoreactor shell permeability for HCF (Fig.~\ref{figrateperm}B) and clearly observe the aforementioned permeability switch below and above the LCST. 
The permeability decreases around one order of magnitude from the swollen to the collapsed state. 
By comparing Figs.~\ref{figrateperm}A and B we clearly see that the nanoreactor shell permeability is the essential ingredient to understand the reaction rate response of nanoreactors in diffusion-controlled reactions. Detailed mesoscopic and microscopic insights on the influence of effective interaction potentials, hydrogel density, and chemistry on the permeability of polymer shells is presented in Sections 3 and 4.

\subsection{Pseudo-unimolecular reactions in nanoreactors}

Bimolecular reactions are typically treated as pseudo-unimolecular when one of the reactants is in large excess with respect to the other. 
The reasoning behind this assumption is that, according to the simple Smoluchowski rate, the reactant in larger concentration would diffuse towards the nanoparticle surface at a much larger rate than the other one. Therefore, when the reactant in limiting concentration arrives to the catalyst, it will always find a reactant of the other species to combine with.
However, this is not always true when considering nanoreactors. In this case, the diffusion rate, \Eq~(\ref{eq:difurateperm}), not only depends on the bulk reactant concentration but also on the shell permeability and thus on the molecular interactions of reactants with the shell. It is thus the combination of both quantities that determines whether a bimolecular reaction can be treated as pseudo-unimolecular or not.

If one of the reactants has a much larger diffusion rate than the other one, {e.g.}, $k_{D^\mathrm{B}}\gg k_{D^\mathrm{A}}$, the total reaction rate, \Eq~(\ref{equa1}), reduces to (see \SItext\ in Ref.~\citenum{Rafa2017})
\begin{equation}\label{eq:totalrate-L2}
k_\mathrm{tot}\rightarrow k_\mathrm{tot}^1
=
\left({k_{D^\mathrm{A}}^{-1}}+k_\mathrm{R}^{-1}\right)^{-1} 
\eqs,
\end{equation}
which is the well-known expression of the total reaction rate in unimolecular reactions, $k_\mathrm{tot}^1$. In this case, the total catalytic time is the sum of the diffusion time of the slower reactant and the surface reaction time. Hence, in nanoreactors, unimolecular reactions can be diffusion- or surface-controlled if $k_D\ll k_\mathrm{R}$ or $k_D\gg k_\mathrm{R}$, respectively. If both rates are comparable in magnitude, the reaction is termed diffusion-influenced. Analogously, a reaction is diffusion- or surface-controlled if $\mathrm{Da}_\mathrm{II}\gg1$ or $\mathrm{Da}_\mathrm{II}\ll1$, respectively, where $\mathrm{Da}_\mathrm{II}=k_\mathrm{tot}/k_D$ is the second Damk\"ohler number.~\cite{Herves:2012fp} 
If both reactants diffuse from the bulk solution, according to \Eq~(\ref{eq:difurateperm}), this condition is satisfied when $c_0^\mathrm{B} \mathcal{P}^\mathrm{B} \gg c_0^\mathrm{A} \mathcal{P}^\mathrm{A}$. This means that one of the reactants should be in a much higher bulk concentration and/or subject to a much larger shell permeability than the other. 

\begin{figure}[t!]
\begin{center}
\includegraphics[width=0.85\linewidth]{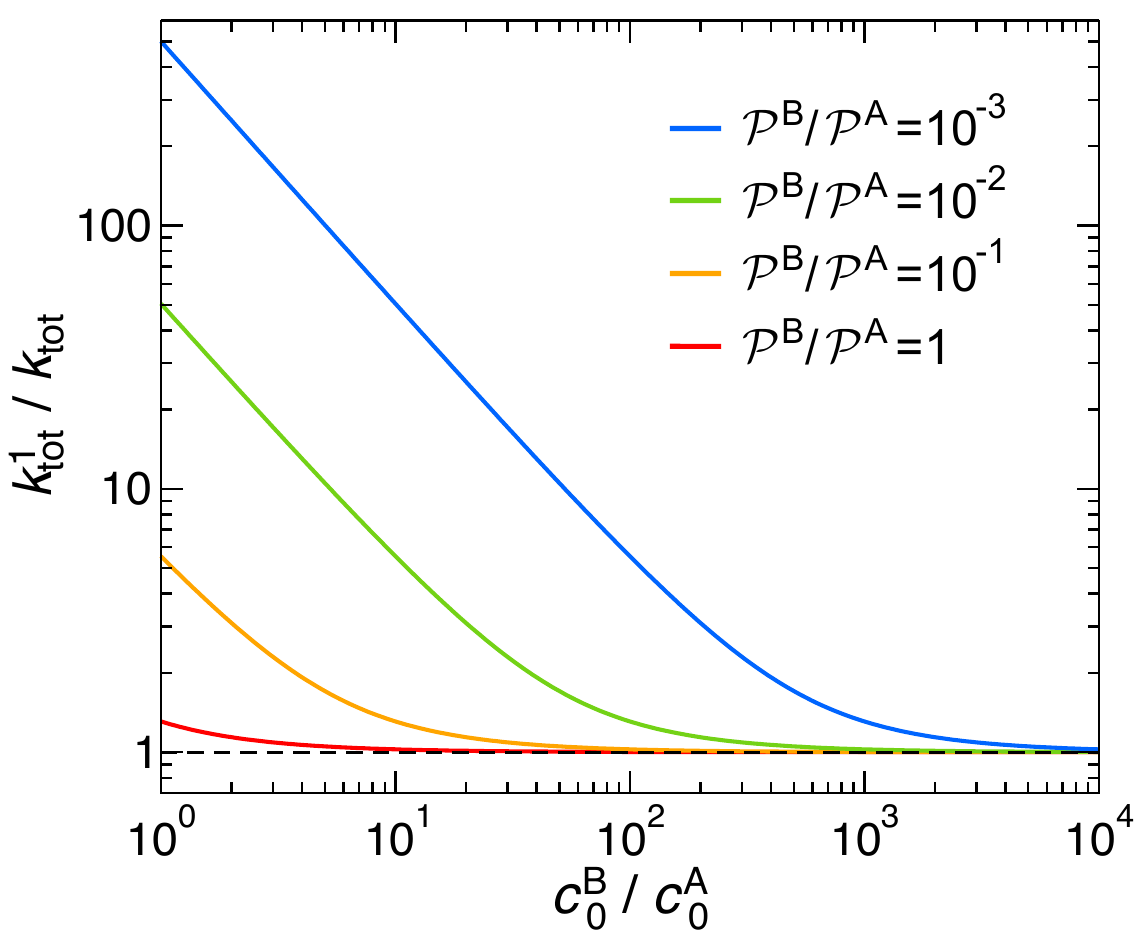}
\caption{Total rate for unimolecular reactions $k_\mathrm{tot}^1$,  \Eq~(\ref{eq:totalrate-L2}), divided by the total reaction rate $k_\mathrm{tot}$ for bimolecular reactions, \Eq~(\ref{equa1}), as a function of the relative reactant bulk concentration, $c_0^\mathrm{B}/c_0^\mathrm{A}$. The  lines stand for different relative nanoreactor shell permeabilities to the reactants, $\mathcal{P}^\mathrm{B}/\mathcal{P}^\mathrm{A}$. We assume $k_\mathrm{R}=k_{D^\mathrm{A}}$.}
\label{figktot1ktot_conc}
\end{center}
\end{figure}

In Fig.~\ref{figktot1ktot_conc} we analyze how large should be the excess of reactant B for the pseudo-unimolecular reaction limit to be valid. This value depends on the relative nanoreactor shell permeability, $\mathcal{P}^\mathrm{B}/\mathcal{P}^\mathrm{A}$.
For simplicity, we consider that the surface rate is equal to the diffusion rate of the reactant in limiting concentration ($k_\mathrm{R}=k_{D^\mathrm{A}}$, diffusion-influenced reaction). When both reactants have the same permeability (red line), the concentration of reactant B should be roughly 10 times larger than the one of A to have a unimolecular reaction.
If we then decrease the shell permeability to the reactant B by 10 times, its concentration has thus to become 100 times higher with respect to that of A to keep this limit.
Figure~\ref{figktot1ktot_conc} also shows that the catalytic rate predicted for a pseudo-unimolecular reaction for the reactant in limiting concentration may differ from the fully bimolecular one by orders of magnitude. Thus, when dealing with nanoreactors, it is necessary to consider not only the difference between the bulk concentrations of the reactants but also the difference in the shell permeability to the reactants.

\begin{figure*}[t!]
\centering
\includegraphics[width=1\textwidth]{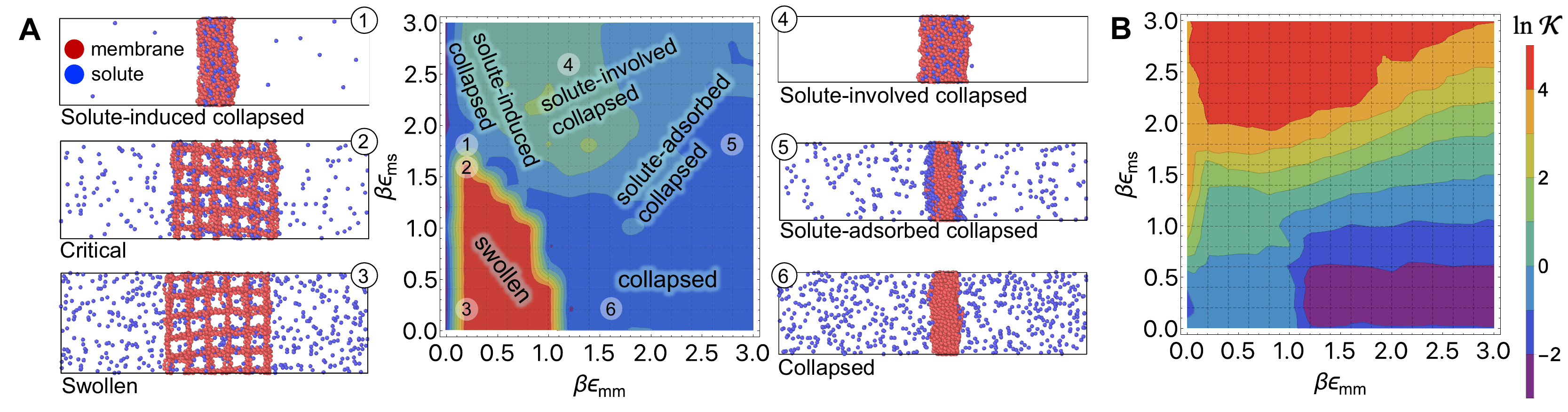}
\caption{
(A)~Various conformational states and regimes in the mesoscopic network membrane--solute system from CG computer simulations are depicted in the main phase diagram (center) depending on the Lennard-Jones (LJ) interactions $\epsilon_\text{mm}$ and  $\epsilon_\text{ms}$: (1) is a `solute-induced collapsed' state, (2) is a `critical' transition line (yellow contour) between the intrinsic (3) `swollen' and (6) `collapsed' states. (4) is a `solute-involved collapsed' state, while (5) is a `solute-adsorbed' collapsed state where solutes adsorb mostly on the membrane surface. For details see text. Reprinted with permission from Ref.~\citenum{kim2017cosolute}, copyright 2017 American Chemical Society.
(B)~Solute partitioning landscape $\mathcal{K}$ depending on $\epsilon_\text{mm}$ and  $\epsilon_\text{ms}$.
}\label{fig:mesofig1}
\end{figure*}

Hence, when considering nanoreactors, care should be taken since in these systems it is not the bulk mobility and concentration that determine the reaction type (diffusion versus surface control), but the values in the polymer shell, which can be the limiting factor. The latter is defined by the shell permeability $\mathcal P$ and can thus strongly differ from the bulk value. Because of the responsive nature of the gating shell of nanoreactors, this dependence crucially implies that the identity of the limiting reactant can switch upon a change in the external stimulus. Failure to recognize this fact can lead to very large discrepancies between the correct and the approximate rate. This theoretical framework for pseudo-unimolecular reactions qualitatively rationalizes the large and sharp variations in catalytic rate observed in the relevant nanoreactor experiments.~\cite{Herves:2012fp,Wu:2012bx,stefano2015,Jia:2016cy}


\section{Partitioning and diffusion:~coarse-grained simulations}

As we have just described, key parameters to understand a nanoreactor's selectivity and rate response to stimuli are the permeability of its polymeric shell and the reactant partitioning within. In the following, we review two selected coarse-grained (CG) simulation studies of partitioning, diffusion, and permeability in model membranes.~\cite{kim2017cosolute, kim2019tuning} Mesoscopic models, neglecting chemical resolution, play a pivotal role not only as a bridge between the aforementioned macroscopic reaction model and the following microscopic all-atom models but also for the deeper understanding of essential physics, e.g., of molecular adsorption and transport in polymer systems. Particle-based simulations on various scales with increasing complexity and chemical detail are now emerging. For the convenience of the reader, we have summarized selected relevant simulation efforts in Tab.~\ref{tab:table2}.

\subsection{Influence of gel volume transition on reactant partitioning in a model polymeric membrane}

Responsive polymers feature a sharp volume transition where the density of the polymer drastically changes. The partitioning of reactants across the volume transition and the feedback of the polymer to the permeation is complex and poorly understood. We thus first discuss a CG simulation model with details described previously~\cite{kim2017cosolute} consisting of permeating reactants in a polymer-based thin membrane (Fig.~\ref{fig:mesofig1}A), where we aim at a qualitative study of the effects of structural transitions of gels on the reactant partitioning and its back-coupling to the volume transition. In the following, we refer to the permeating reactants generally as `solutes'.   

The membrane is constructed as cross-linked semi-flexible network of polymers formed on a regular cubic lattice, \cite{Aydt2000,Lu2002,pnetz1997,Erbas2015,Erbas2016,Li2016} and the solutes can diffuse throughout the membrane and the bulk regions.
This enables a direct sampling of the solute partitioning from the simulations simply according to \Eq~(\ref{eq:Kdirect}).  
We use the Lennard-Jones (LJ) pair potential and its size unit $\sigma$ as the diameter for all particles and the monomer--monomer (bonded) distance in the polymers.   
In order to model such a gel in the presence of various solutes, we employ inter- and intra-particle interactions in terms of LJ pair potentials. We focus on two key interaction parameters: The membrane--membrane interaction $\epsilon_\text{mm}$ controls the solvent quality, turning it from good to poor upon the increase of $\epsilon_\text{mm}$. The membrane--solute interaction $\epsilon_\text{ms}$ governs the membrane--solute {\it coupling} and thus models different kinds of solutes. For the solute--solute interaction we always use $\epsilon_\text{ss}=0.1~k_\text{B}T$, essentially being a steep ($r^{-12}$) repulsion. 

\begin{figure*}[t]
\centering
\includegraphics[width=1\textwidth]{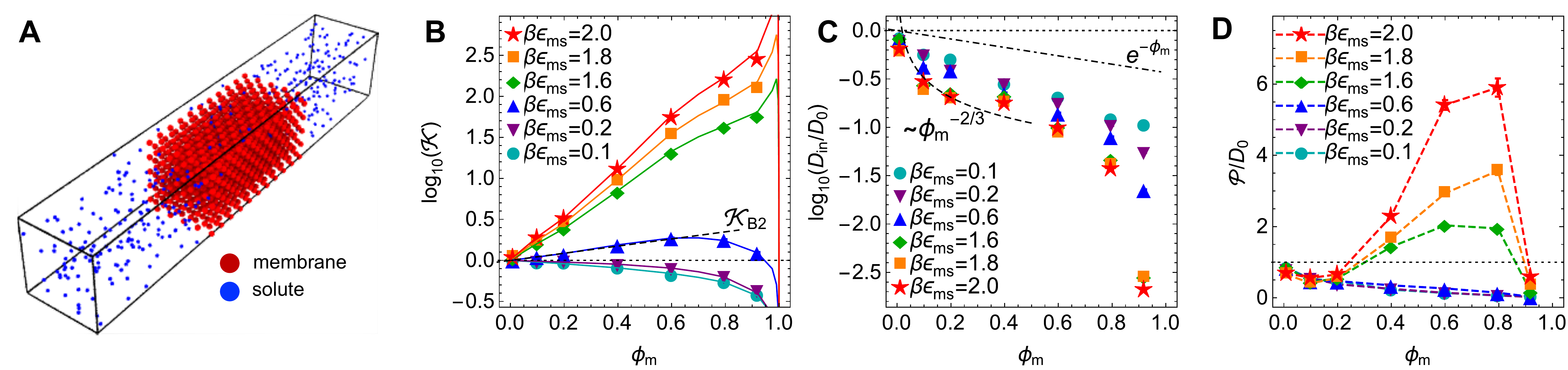}
\caption{
(A)~Snapshot of the mesoscopic lattice membrane--solute system. The membrane sites (red) are fixed on a face-centered-cubic (fcc) lattice with the volume fraction $\phi_\text{m}$, and the penetrating solutes (blue) are diffusing and interacting via the LJ potentials with $\epsilon_\text{ms}$. 
(B)~Solute partitioning $\mathcal{K}(\phi_\text{m})$ at different $\epsilon_\text{ms}$. The solid lines depict the exact relation, \Eq~\eqref{eq:K1}, and the dashed line depicts the approximated partitioning $\mathcal{K}_\text{B2}$ with $\beta \epsilon_\text{ms} = 0.6$ (see text for details).
(C)~Solute diffusivity $D_\text{in}(\phi_\text{m}) / D_0$ with different $\epsilon_\text{ms}$. The dashed lines depict the approximation $D_\text{in} / D_0 = \exp ( - \phi_\text{m} )$ valid for low $\epsilon_\text{ms}$ and $\phi_\text{m}$, and the scaling $D_\text{in} / D_0 \sim \phi_\text{m}^{-2/3}$ valid for high $\epsilon_\text{ms}$ and low $\phi_\text{m}$.
(D)~Permeability $\mathcal{P}(\phi_\text{m})/D_{0}$ at different $\epsilon_\text{ms}$.
Reprinted with permission from Ref.~\citenum{kim2019tuning}, copyright 2019 American Physical Society.
}\label{fig:mesofig}
\end{figure*}

In Fig.~\ref{fig:mesofig1}A we show the landscape of the gel structural phases, depending on both interaction parameters. The red region depicts swollen states, where the gel volume is relatively large, while the blue region indicates collapsed states. Without the solutes ($\epsilon_\text{ms}=0$) our model exhibits a collapse transition at around $\epsilon_\text{mm}\simeq 1.0~k_\text{B}T$. In the presence of the solutes, however, the picture becomes more complex: 
The `critical' transition line (yellow contour line) between swollen and collapsed states depends substantially on the membrane--solute interaction, as shown by the label (2). The stark color contrast around this critical line signifies the sharp transition. 
In addition, one can identify in total five distinct phase regions (or states), classified into (1) ``solute-induced collapsed'', (3) ``swollen'', (4) ``solute-involved collapsed'', (5) ``solute-adsorbed collapsed", and (6) ``collapsed'' states, indexed by the numbers in the colored center panel of Fig.~\ref{fig:mesofig1}A.  
Interestingly, the ``solute-induced collapsed'' state (state 1) can occur even in good solvent conditions, where the membrane undergoes a relatively sharp collapse transition induced by a strong `bridging' attraction between the solutes and the network monomers.  The effect has been reported in computer simulations before but only on the single polymer level.~\cite{ Heyda2013, mukherji2013coil, mukherji2014polymer, rodriguez2015mechanism,  rodriguez2015urea,rika1990intermolecular, lee1997effects, heyda2017guanidinium}
The ``solute-involved'' collapsed state (state 4) occurs at the intermediate solvent quality  where  the membrane collapses with most of the internal solutes embedded, yielding a bulkier collapsed gel than the intrinsically collapsed case. 
The ``solute-adsorbed collapsed'' state (state 5) is an example for the limiting case, where both of the membrane--membrane and membrane--solute attractions are strong, but the first one dominates and excludes the solutes,  therefore leading to a strong surface accumulation of those.

The solute partitioning $\mathcal{K}$ (on a log-scale) averaged over the membrane slab is shown in Fig.~\ref{fig:mesofig1}B in a 2D-landscape plot, and is related to the transfer free energy from the bulk into the network, $\Delta G = - k_\text{B} T \ln \mathcal{K}$, which quantifies the average transfer free energy for the solute transfer from bulk to the membrane. The partitioning varies by several orders of magnitude, depending not only on the membrane--solute interaction but also significantly on the solvent quality.  The partitioning overall becomes large (i.e., higher adsorption) as the membrane--solute attraction, $\epsilon_{\rm ms}$, increases, while it has large regions of unity in the swollen states (light blue-greenish areas).  Note that when compared with the structural landscape in Fig.~\ref{fig:mesofig1}A, both extrema of $\mathcal{K}$ (i.e., the minimum and the maximum) are in the collapsed regions, indicating that the collapsed phase can relate to extremely different partitionings and there is no unique mapping.  Moreover, at intermediate values of $\epsilon_{\rm ms}$, $\mathcal{K}$ is a nonmonotonic function of the solvent quality, meaning that it can be maximized by an optimal solvent quality.  The maximization of the partitioning is in fact a quite general feature in attractive but crowded systems as we will discuss in the following section.

To sum up, the CG simulation model of a polymer network in the presence of solutes reveals a rich topology of structure phases and their relation to solute partitioning, entering the rate equations  \Eqs~(\ref{eq:surfacerate}) and~(\ref{eq:kDkD0}). In particular, for very attractive solute--membrane interactions ($\epsilon_\text{ms} \gtrsim 1~k_\text{B}T$) the network structure and partitioning are coupled. 
The mesoscopic model thus provides a landscape of the partitioning, thereby bridging the macroscopic continuum model and microscopic discrete data in terms of the generic interaction parameters. In addition, the results will be helpful for the interpretation of experiments for certain polymer--reactant systems and could also be useful to design {\it feedback-systems} where the local reactant (or product) concentration may couple back to polymer structure in a prescribed way.

  
\subsection{Partitioning, diffusion, and permeability in a model lattice membrane}

Now we present a related but different CG model~\cite{kim2019tuning} of a membrane--solute system (Fig.~\ref{fig:mesofig}A) in order to study permeability in dense media qualitatively on a generic level. We demonstrate how the permeability can be tuned massively in magnitude by systematically varying the membrane--solute interactions and the density of the membrane. The study also gives important insights about how partitioning and diffusion are correlated.

The permeability is defined following the solution--diffusion theory~\cite{yasuda1969permeability3,robeson, Baker, gehrke, chauhan,Thomas2001transport,Ulbricht2006, Baker2014,park2017} by \Eq~(\ref{eq:P}).
There have been pioneering theoretical models to elucidate the transport phenomena in membranes  ~\cite{Yasuda1968,yasuda1969permeability2,yasuda1969permeability3,robeson, Baker, gehrke, masaro1999physical, Amsden1998} based on simple theories for either partitioning or diffusion.  Recently, a simulation study revealed maximization of partitioning of penetrating solutes in polymer membranes tuned by the polymer volume fraction.~\cite{monchoPCCP2018} Diffusion in dense membranes is usually quite complex and highly dependent on details of the interaction potentials.~\cite{lyd,obliger, cai2015hopping, Ben2, zhang2017molecular,  zhang2017correlated,Huskey2016, kanduc2018diffusion} Nevertheless, there have been no comprehensive studies on the permeability $\mathcal{P}$, being a product of partitioning and diffusion.

In the CG model membrane--solute system as shown in Fig.~\ref{fig:mesofig}A the diffusive solutes are ideal ($\epsilon_\text{ss}=\sigma_\text{ss}=0$), and the membrane consists of immobile interaction sites,  located on a face-centered-cubic (fcc) lattice with a fixed unit cell size $l$, variation of which tunes the monomer packing fraction $\phi_\text{m}$. The simplicity of such an ordered and rigid model membrane with ideal solutes renders the problem easier for interpretation and perhaps theoretically tractable. 
The ideal solutes diffuse throughout the simulation box but interact only with the membrane sites via the LJ potential with the coupling strength $\epsilon_\text{ms}$. 
For the ideal solutes the partitioning can be exactly computed via the transfer free energy shown in \Eq~(\ref{eq:partitioning}). In our case $\Delta G = -\kB T\, \ln\, \overline{ \rme^{-\beta H_\text{ms}} }$~\cite{Leo1971}, where 
$H_\text{ms}(\mathbf r)=\sum_i U_{\rm ms}(|\mathbf r - \mathbf r_i|)$ is the total Hamiltonian (summing over all membrane sites $i$), and $\overline{x} \equiv \int \text{d}V x / V_\text{m}$ is the volume average, yielding
\begin{equation}
\label{eq:K1} \mathcal{K}  =  \overline{ \rme^{-\beta H_\text{ms}} }.
\end{equation}
which verifies the simulation results (Fig.~\ref{fig:mesofig}B).

The computed partitioning as a function of the membrane volume fraction $\phi_\text{m}$ is shown in Fig.~\ref{fig:mesofig}B for various $\epsilon_\text{ms}$. 
For relatively low membrane--solute couplings ($\beta\epsilon_{\rm ms} \lesssim 0.3$), the LJ interaction between solutes and membrane sites is essentially repulsive (signified by a positive second virial coefficient), and the partitioning monotonically decreases as the membrane becomes dense, owing to the dominant exclusion by the membrane. For intermediate couplings around $\beta\epsilon_\text{ms} = 0.6$, which is moderately attractive, partitioning reaches a maximum at an optimal membrane density around $\phi_\text{m} = 0.6$. The partitioning maximization is attributed to a balance between adsorption and steric exclusion.  
In addition, a leading order approximation of $\mathcal{K}$ on a two-body level, $\mathcal{K}_\text{B2}(\phi_\text{m},\epsilon_\text{ms})=\exp\left[-2 c_\text{m} B_2^\text{ms}\right]$ for $\beta \epsilon_{\rm ms}=0.6$ is depicted by the dashed line, where $c_{\rm m}\propto \phi_{\rm m}$ is the membrane concentration, and $B_2^\text{ms}$ is the second virial coefficient.
Figure~\ref{fig:mesofig}C shows the solute diffusivity $D_\text{in} / D_0$ in the membrane as a function of $\phi_\text{m}$, rescaled by the free diffusivity in the bulk.  As the membrane becomes dense, $D_\text{in}$ tends to exponentially decrease, showing more complex behavior with higher couplings. We compare the simulation results with scaling theories for diffusion in two limiting cases. 
The upper dashed line indicates the limiting law $D_\text{in} / D_0 = \exp (- \phi_\text{m} )$ based on the `volume-exclusion' theory.~\cite{HAUS1987,masaro1999physical, Amsden1998, Ghosh2014a, lyd}
It is indeed valid only for low couplings, which acts essentially repulsive.
For high membrane--solute attractions and low membrane density, the diffusivity follows the power law $D_\text{in} / D_0 \sim \phi_\text{m}^{-2/3}$. The scaling relation is derived by the limiting law from the Kramers' barrier crossing over the distance $l\sim\phi^{-1/3}$, and therefore $D_{\rm in} \sim l^2/\tau \sim \phi_\text{m}^{-2/3}$.~\cite{kim2019tuning} 

The resulting permeability, the product of $\mathcal{K}$ and $D_{\rm in}$, shown in Fig.~\ref{fig:mesofig}D, exhibits intriguing features.
For essentially repulsive solutes, $\mathcal P$ decreases monotonically as the membrane density increases, and the overall magnitude is below unity, almost approaching zero for very dense membranes.  We speculate that this essentially repulsive case may be the scenario in the experiments with the highly charged reactant HCF in Section 2.1, which probably does not like to enter the collapsed gel, but this suspicion needs further scrutiny. On the other hand, for high couplings (attraction) the permeability is first minimal around $\phi_\text{m} = 0.1$, then maximized at large membrane densities $\phi_\text{m} \simeq 0.8$. The permeability vanishes at the maximum overlapping density ($\phi_\text{m} \sim 1$), where no percolating holes for diffusion are present anymore. This demonstrates a clear maximization of permeability when the system is highly attractive and dense.  The nonmonotonic behavior of permeability results from drastic nontrivial cancellations between the partitioning and the diffusivity, which exponentially increase and decrease, respectively. The massive cancellation between two largely varying functions over several orders of magnitude yields a permeability of the order of unity,~\cite{kim2019tuning} implying a high potential for fine-tuning of the permeability behavior in experiments by small changes in density or interactions.

Mesoscopic models of membrane--solute systems demonstrate that the permeability, typically resulting from large cancellation effects of partitioning and diffusion, is very sensitively tuned by the effective interaction potentials and the membrane density. The results indicate that most drastic selectiviy effects are at high membrane densities and significant ($\gtrsim k_\text{B}T$) membrane--solute attractions. The effective potentials in realistic material design assemblies can be somewhat controlled by various external parameters, such as temperature, ionic strength, pH, and possibly various additives in the solution. The results from the mesoscopic models thus provide useful physical insight and may bear important applications in design and engineering of molecular systems to achieve a selective transport by fine-tuning interactions and topologies, particularly in highly attractive membrane systems.


\begin{table*}[t!]
\small
  \begin{center}
    \caption{Survey of selected publications on computer simulations of diffusion $D$, partitioning $\mathcal K$, permeability $\mathcal P$, or related adsorption or transport phenomena of (co)solutes in polymers. Abbreviations: All-atom (AA), Coarse-Grained (CG), Molecular Dynamics (MD),  Monte-Carlo (MC), Langevin Dynamics (LD), Brownian Dynamics (BD), Dissipative Particle Dynamics (DPD).}
    \label{tab:table2}
\resizebox{\textwidth}{!}{
    \begin{tabular}{|p{1.1cm}|p{2.0cm}|p{2.5cm}|p{2.7cm}|p{3.3cm}|p{9cm}|}
      \hline
      Ref. & Simulation methods & Architecture & Polymer & Resulting quantity & Comment \\
      \hline
            \citen{wu2009effect} & AA-MD & swollen cross-linked network & polyethyleneglycole & diffusivity of water, ions, rhodamine & water content: 75--91\%, mesh size: 2.3--5.5~nm \\
      \citen{borjesson2013molecular} & AA-MD& collapsed & polyethylene & diffusivity and partitioning of oxygen and water & permeability for oxygen 5--6 orders of magnitude larger than for water \\
      \citen{karlsson2004molecular} & AA-MD & collapsed & poly(vinyl alcohol) & diffusivity of O$_2$& water uniformly distributes \\
            \citen{muller1998molecular,muller1998diffusion} & AA-MD & collapsed & poly(vinyl alcohol) & diffusivity of water & hydrogel with 4--40~\% water \\
      \citen{fritz1997molecular} & AA-MD & collapsed & polydimethylsiloxane & diffusivity of water and ethanol & water/ethanol mixtures; water molecules faster than ethanol \\
      \citen{kucukpinar2003molecular} & AA-MD~+~transition state approach & collapsed & polystyrene and its copolymers & diffusivity and partitioning of gas and water molecules & \\
               \citen{rodriguez2014direct} & AA-MD & single chain & PNIPAM & adsorption of urea & studying volume phase transition \\
            \citen{schroer2016stabilizing} & AA-MD & single chain & PNIPAM & adsorption of TMAO, urea & studying volume phase transition \\
      \citen{adroher2017conformation} & AA-MD & collapsed and solvent phase & PNIPAM & partitioning of ions & thin core-shell membrane, direct measuring of partitioning\\
      \citen{deshmukh2009molecular} & AA-MD & cross-linked network & PNIPAM & diffusivity of water; volume transition & studying volume transition; cross-linking inhibits the collapse \\
      \citen{botan2016direct} & AA-MD & collapsed and solvent phase & PNIPAM (3mer) & water--polymer coexistence & also 30mer of PNIPAM: no conclusions on chain configuration \\
      \citen{perez2015anions} & AA-MD & swollen \& collapsed finite aggregate & PNIPAM & partitioning of large ions & umbrella sampling of the potential of mean force of the ions \\
      \citen{aydt2000swelling} & Gibbs-ensemble CG-MD & cubic network & bead-spring & solvent sorption and swelling isotherm & effects of solvents on swelling \\
      \citen{Lu2002} & Two-box--particle-transfer CG-MD & cubic network & bead-spring & solvent sorption and swelling isotherm & effects of cross-linkers on swelling \\
      \citen{Escobedo1997,Escobedo1999} & CG-MC & tetra-functional network & bead-spring & solvent sorption and swelling isotherm & effects of polymer network density and deformation on swelling \\
      \citen{pnetz1997} & CG-MC & cubic network & rigid rod & cosolute diffusivity & effects of cosolute size and polymer density on cosolute diffusivity \\
      \citen{Erbas2015} & CG-LD & cubic network & charged bead-spring & energy conversion & effects of compression and solvents on energy contribution \\
      \citen{Erbas2016} & CG-LD & highly swollen cubic network & charged bead-spring & adsorption and conformational response & counterion-induced deformation \\
      \citen{Li2016} & LD & cubic network & charged bead-spring & ion transport & effects of electrostatic coupling between polymer and ions on ion transport \\
      \citen{monchoPCCP2018} & CG-MC & tetra-functional network & bead-spring & cosolute partitioning & effects of polymer density on partitioning \\
      \citen{zhang2017molecular} & CG-MD & polymer melt & semi-flexible & gas partitioning, diffusivity and permeability & effect of gas size and polymer semi-flexibility on gas transport \\
      \citen{sandrin2016diffusion} & CG-BD & cubic network & bead-spring & cosolute diffusivity & effects of cosolute density on cosolute diffusivity \\
      \citen{masoud2010permeability} & CG-DPD & random network & semi-flexible & permeability and cosolute diffusivity & effects of porosity and deformation on permeability \\
      \citen{hansing2018hydrodynamic,hansing2018particleBJ,hansing2018particle} & CG-BD & cubic network & rigid rod & cosolute diffusivity & effects of interactions, hydrodynamics, and network heterogeneity on cosolute diffusivity \\
      \citen{zhou2009brownian} & CG-BD & random cubic network & rigid rod & cosolute diffusivity & effects of network porosity, flexibility, degree of cross linking, and electrostatic interaction on cosolute diffusivity \\
      \hline
    \end{tabular}
    }
  \end{center}
\end{table*}

\section{Partitioning and diffusion:~all-atom simulations}

The advantage of the mesoscopic simulations in the previous section is that we can obtain fundamental and qualitative insights on how permeability depends on basic input parameters such as interaction energies, lattice geometry, and single solute diffusion. However, in experiments we deal with specific, chemical systems, where the effects of interactions are highly convoluted and solvation, polarity, electrostatics, and specific steric constraints come explicitly into play. Hence, for a more detailed insight and quantitative numbers for the continuum approach to reaction rates in Section~2, we need to resort to higher resolution, molecular dynamics computer simulations. All-atom simulation studies of partitioning and diffusion through polymer networks with increasing complexity and chemical detail are now emerging and growing in the literature. Selected works in this field are given in Tab.~\ref{tab:table2}. In the following, we will review our recent efforts to understand partitioning and diffusion of solutes  in  swollen and collapsed PNIPAM hydrogels by all-atom (AA) MD simulations.~\cite{kanduc2017selective, milster2019crosslinker, kanduc2018diffusion, kanduc2019free} 

\begin{figure*}[h]\begin{center}
\begin{minipage}[b]{0.85\textwidth}\begin{center}
\includegraphics[width=\textwidth]{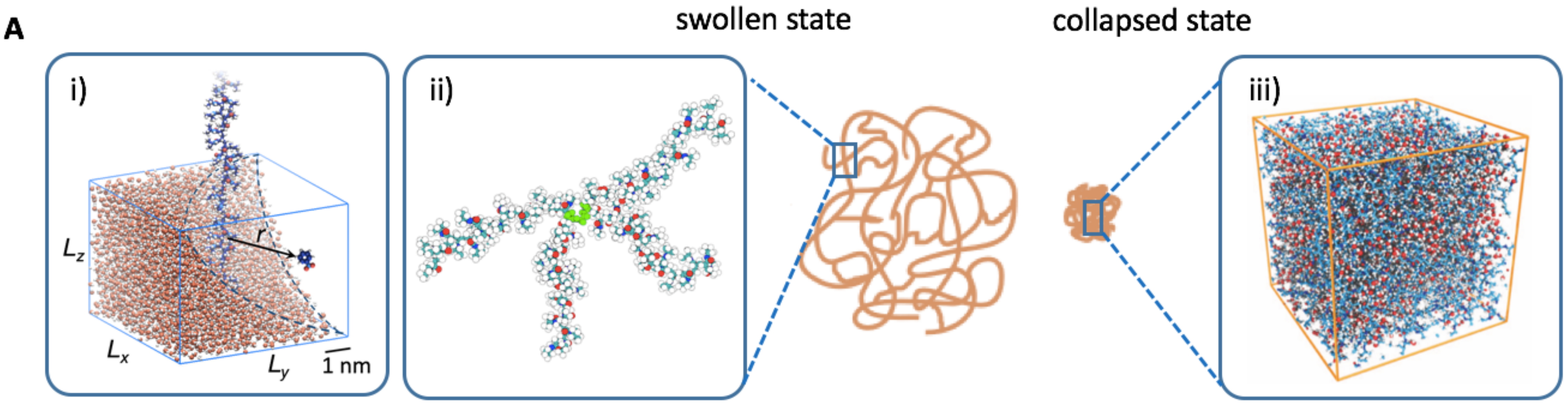}
\end{center}\end{minipage}
\begin{minipage}[b]{0.79\textwidth}\begin{center}
\includegraphics[width=\textwidth]{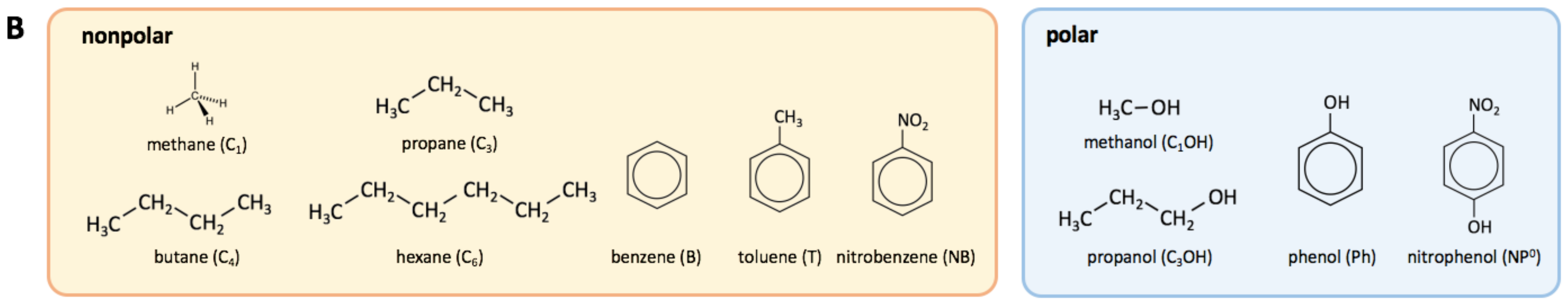}
\end{center}\end{minipage}
\caption{(A) Atomistic modeling of PNIPAM hydrogels: i) elongated, infinitely long chain (mimicking a part of a swollen network where the adjacent chains are far apart), ii) a cross-linker connected with four chains in a tetrahedral structure (representing a unit of a swollen network), iii) dense aggregate of PNIPAM polymers at 340~K (a model for a collapsed PNIPAM hydrogel).
(B)~Solute molecules in our study; polarity is characterized by the hydroxyl (OH) group.}
\label{fig:AA-snaps}
\end{center}\end{figure*}


\subsection{Swollen state}
\begin{figure*}[t!]\begin{center}
\begin{minipage}[b]{0.24\textwidth}\begin{center}
\includegraphics[width=\textwidth]{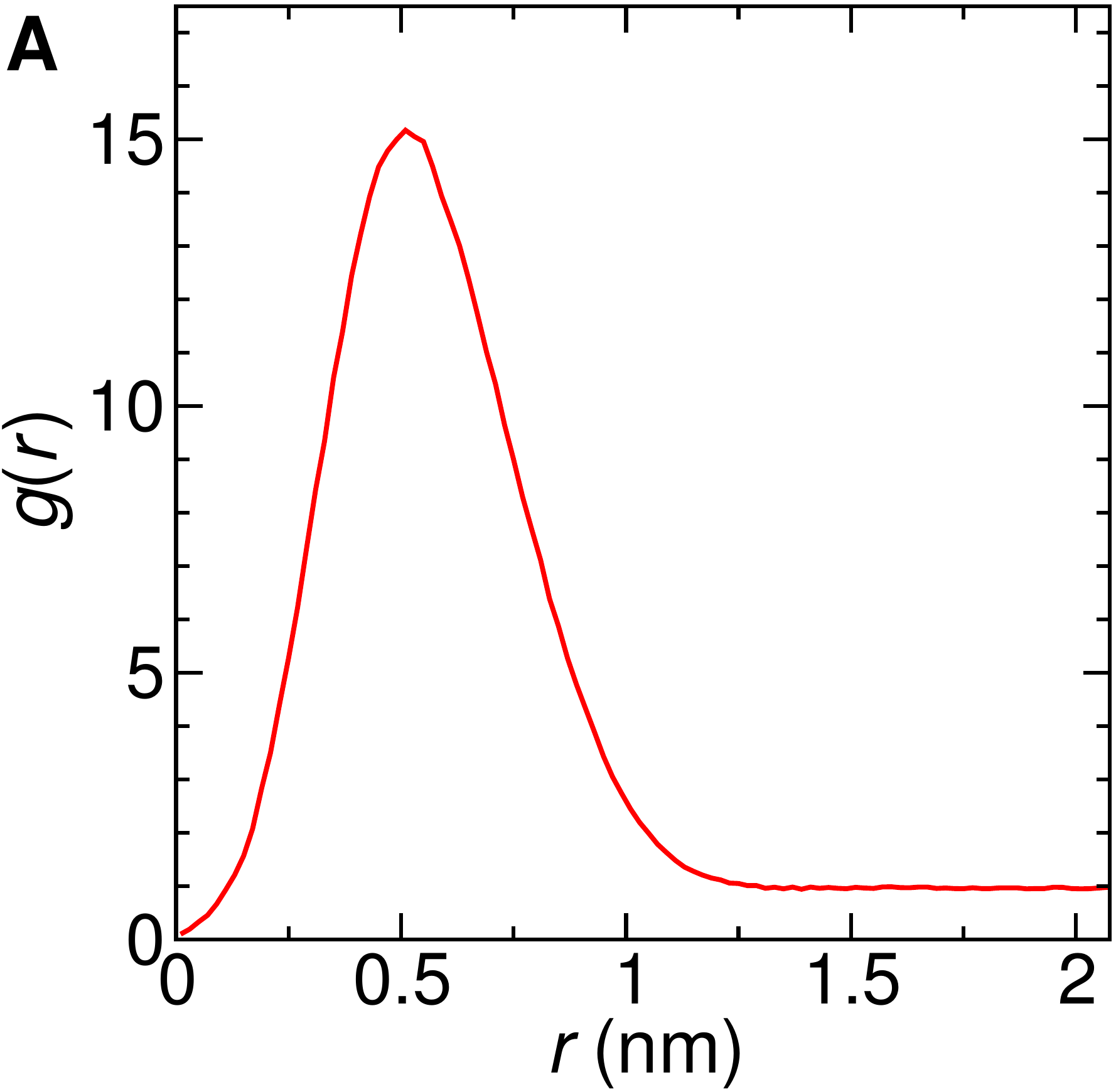}\vspace{-1ex}
\end{center}\end{minipage}\hspace{2ex}
\begin{minipage}[b]{0.62\textwidth}\begin{center}
\includegraphics[width=\textwidth]{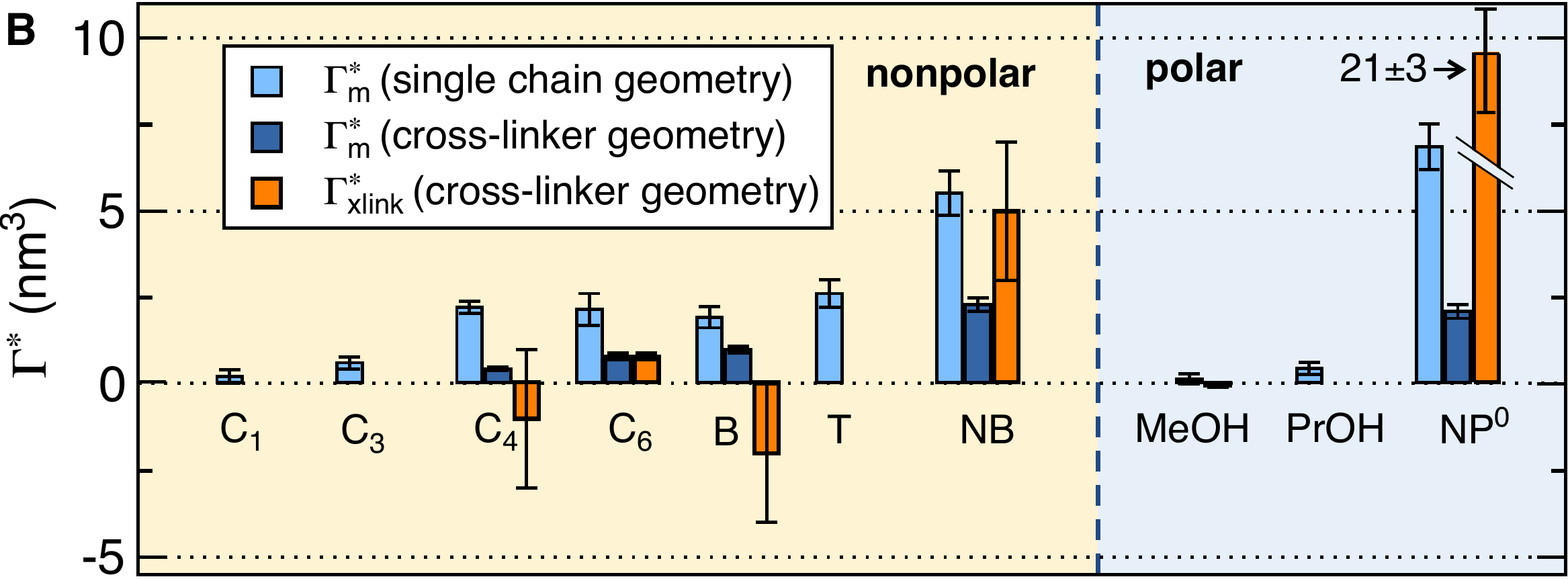}
\end{center}\end{minipage} 
\caption{(A) Cylindrical RDF of backbone--NB for an extended PNIPAM chain~\cite{kanduc2017selective} (see \Fig~\ref{fig:AA-snaps}A.i).
(B)~Adsorption coefficients  of various solutes (see \Fig~\ref{fig:AA-snaps}B) to a PNIPAM monomer $\Gamma^*_\trm{m}$ (from single-chain geometry and the OPLS~\cite{kanduc2017selective} force field, and from cross-linker geometry using the OPLS-QM2~\cite{milster2019crosslinker} force field) and to a cross-linker $\Gamma^*_\trm{xlink}$~\cite{milster2019crosslinker} (OPLS-QM2 force field) at 300 K.}
\label{fig:molecules}
\end{center}\end{figure*}

In order to model the swollen state of a hydrogel shell one can focus on one elongated PNIPAM chain, as shown in \Fig~\ref{fig:AA-snaps}A.i, where the chain is replicated through periodic boundary conditions. The cylindrical geometry allows for a simple extraction of adsorption properties~\cite{kanduc2017selective} of solvated molecules in the solution.
The first step is to evaluate the cylindrical radial distribution function (RDF) of the solute molecules from the backbone, $g_\trm{2D}(r)$, as shown in an example for nitrobenzene (NB) in \Fig~\ref{fig:molecules}A.
The adsorption coefficient $\Gamma_\trm{m}^*$ per monomer of the polymer is then obtained by integration along the spatial coordinates,~\cite{horinekJPCA2011}
\begin{equation}
\Gamma_\trm{m}^*=\Delta L_\trm{m} \int_0^\infty [g_\trm{2D}(r)-1] 2\pi r\rmd r,
\end{equation}
where $\Delta L_\trm{m}=0.265$~nm is the distance between neighboring monomers in the chain.
The total adsorbed number of molecules $\Gamma_\trm{chain}$ on the chain is proportional to the number of monomers $N_\trm{m}$ and, in the infinite dilution limit, to the bulk solute concentration $c_0$, 
\begin{equation}
\Gamma_\trm{chain}=\Gamma_\trm{m}^* N_\trm{m} c_0,
\label{eq:GammaTOT}
\end{equation}

Another setup of swollen hydrogels, shown in \Fig~\ref{fig:AA-snaps}.ii, mimics the cross-linker unit of a hydrogel network, and thus lends itself to study the influence of cross-linkers on adsorption of molecules. 
In our previous study~\cite{milster2019crosslinker}, we considered a very common N,N'-methylenebisacrylamide (BIS) cross-linker, connecting four PNIPAM chains with their ends tethered in a tetrahedral geometry. 
The solute molecules in general adsorb in different proportions to the chain regions and the cross-linker neighborhood.
The overall adsorption in the radial interval $[r_1,r_2]$ from the cross-linker is obtained in a straightforward manner by integrating the (spherical) RDF $g(r)$ of the solutes
\begin{equation}
\Gamma(r_1,r_2)=c_0 \int_{r_1}^{r_2} [g(r)-1] 4\pi r^2\rmd r.
\label{eq:Gamma3D}
\end{equation}
With this, we can scan the adsorption in different regions with respect to the cross-linker. It also allows us to evaluate $\Gamma_\trm{m}^*$, as in the single-chain geometry. Finally, the total adsorption can be deconvoluted  into two contributions,
\begin{equation}
\Gamma_\trm{tot}=\Gamma_\trm{chain}+\Gamma_\trm{xlink}.
\label{eq:GammaCX}
\end{equation}
The adsorption on the chains $\Gamma_\trm{chain}$ (unperturbed by the presence of cross-linker) is given by \Eq~(\ref{eq:GammaTOT}), whereas $\Gamma_\trm{xlink}$ represents the effect on the adsorption due to the presence of the cross-linker. The value of $\Gamma_\trm{xlink}$ can be evaluated from known $\Gamma_\trm{tot}$ and $\Gamma_\trm{chain}$. In the infinite-dilution limit, the adsorption on the cross-linker is proportional to the bulk solute concentration, $\Gamma_\trm{xlink}=\Gamma_\trm{xlink}^* c_0$, where $\Gamma_\trm{xlink}^*$ is the adsorption coefficient of the cross-linker.

The resulting adsorption coefficients $\Gamma_\trm{m}^*$ are shown in \Fig~\ref{fig:molecules}B (blue shaded bars), from the single-chain~\cite{kanduc2017selective} and cross-linker~\cite{milster2019crosslinker} geometries. Quite generally, the adsorption grows with the molecular size. 
The effect of the cross-linker, $\Gamma_\trm{xlink}^*$, is shown in \Fig~\ref{fig:molecules}B by orange bars: 
 The apolar compounds C$_4$, C$_6$, and B show a low affinity to the cross-linker. In contrast, the adsorption of nitro-aromatic solutes to the cross-linker is significant, in particular for NP$^0$.~\cite{milster2019crosslinker} NB shows more than doubled and NP$^0$ even an order of magnitude higher adsorption to the cross-linker region than to a monomer of the polymer. 
 Note that the BIS cross-linker has two amide groups  and is slightly more hydrophilic than the PNIPAM chain, hence favoring polar molecules.
 
\begin{figure}\begin{center}
\begin{minipage}[b]{0.3\textwidth}\begin{center}
\includegraphics[width=\textwidth]{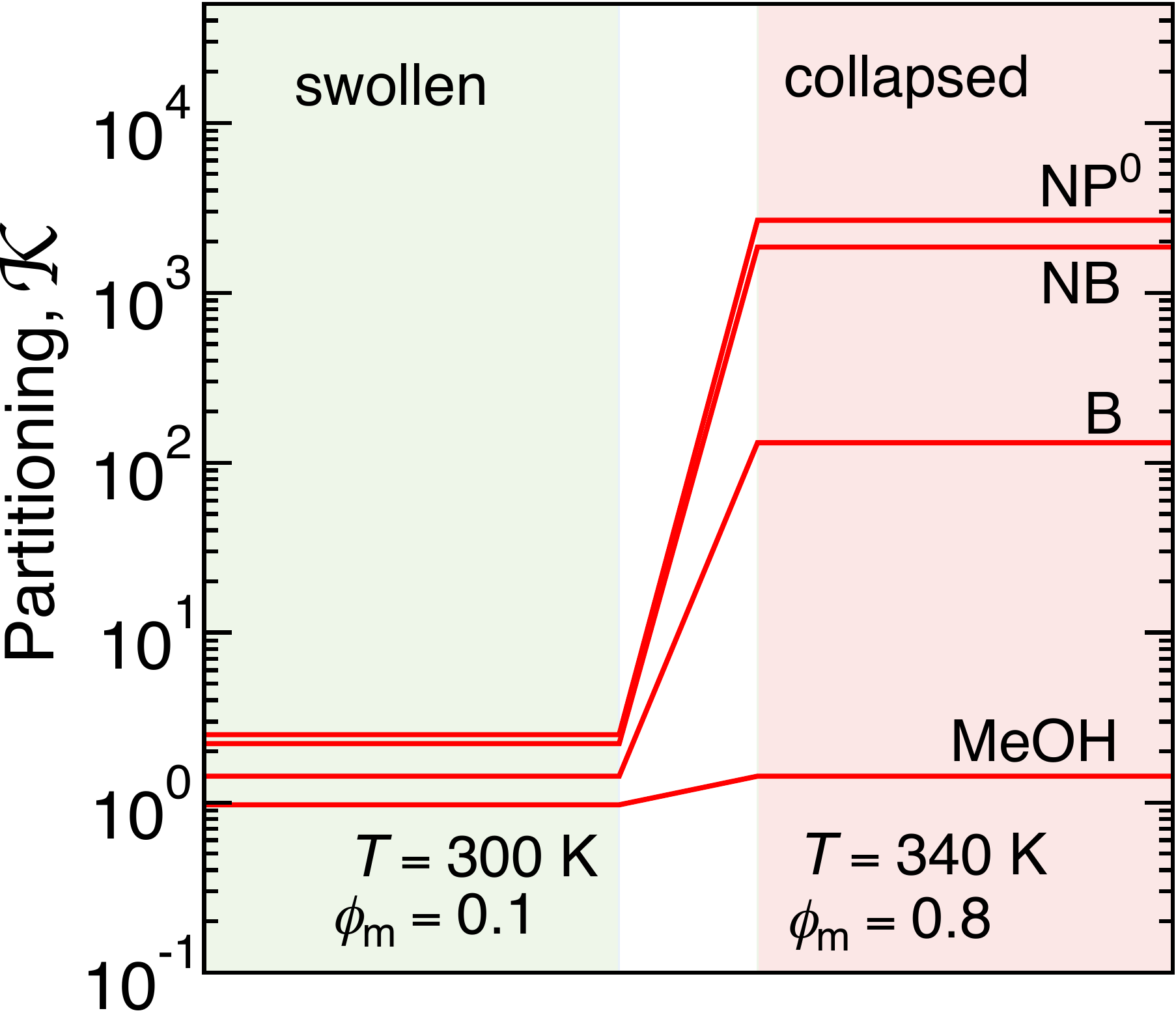}
\end{center}\end{minipage}
\caption{Partitioning of several molecules resulting from the atomistic models of a swollen (at 300~K and polymer fraction of $\phi_\trm{m}=0.1$) and collapsed (at 340~K and $\phi_\trm{m}=0.8$) state of a PNIPAM gel.
The values for the swollen state are computed from \Eq~(\ref{eq:Kswollen}) and assuming  polymer volume fraction $\phi_\trm{m}=0.1$, whereas the values for the collapsed state are computed from \Eq~(\ref{eq:Kdirect}).
}
\label{fig:AA-K}
\end{center}\end{figure}

From the known adsorptions on individual chains and cross-linkers we can predict the partitioning in extensive hypothetical swollen polymer architectures, such as hydrogels.
The partitioning follows from $\mathcal{K}=1+\Gamma_\trm{tot}/(c_0 V)$, where $V$ is the volume of the gel and $\Gamma_\trm{tot}$ the total adsorption of molecules on all the chains and cross-linkers [\Eq~(\ref{eq:GammaCX})], which leads to
\begin{equation}
\mathcal{K}=1+n_\trm{m}\Gamma_\trm{m}^*+n_\trm{xlink}\Gamma_\trm{xlink}^*,
\label{eq:Kswollen}
\end{equation}
where $n_\trm{m}$ and $n_\trm{xlink}$ are the monomer and cross-linker number densities, respectively.
The  former can be easily linked to the polymer volume fraction $\phi_\trm{m}$ as $n_\trm{m}=\phi_\trm{m}/(\pi R_0^2 \Delta L_\trm{m})$, where $R_0=0.5$~nm is an estimated effective radius of the polymer chain.~\cite{kanduc2017selective}
Assuming $\phi_\trm{m}=0.1$ for a typical architecture of a swollen state, we compute $\mathcal{K}$ for several solutes from the obtained MD parameters in \Fig~\ref{fig:AA-K}. The values range around unity, $\mathcal{K}\approx$1--3, as also resulted from the CG models in \Fig~\ref{fig:mesofig}B for this polymer fraction range.
As we will see in the following, the collapsed state can give rise to much higher partitioning.

\subsection{Collapsed state}

The collapsed state of the PNIPAM hydrogel can be modeled as a bulk of aggregated polymeric chains (in our case 20 monomeric units long) at 340~K (above the LCST), where cross-linkers are ignored. 
The amount of sorbed water between the polymeric chains is chosen such that it corresponds to the chemical equilibrium with bulk water.~\cite{kanduc2018diffusion}  The amount of water in the collapsed state depends on temperature, and amounts to around 20~wt.~\% (somehow less than 
 experimental estimates of around 30~wt.~\%~\cite{dong1990synthesis, sasaki1999dielectric,raccis2011probing, kaneko1995temperature, dong1986thermally}), which roughly correspond to the polymer volume fraction of $\phi_\trm{m}=0.8$. Note that this is in the range of packing fractions for which we observed the most interesting behavior of permeability in the CG simulations in Section~3.

\begin{figure*}[t!]\begin{center}
\begin{minipage}[b]{0.14\textwidth}\begin{center}
\includegraphics[width=\textwidth]{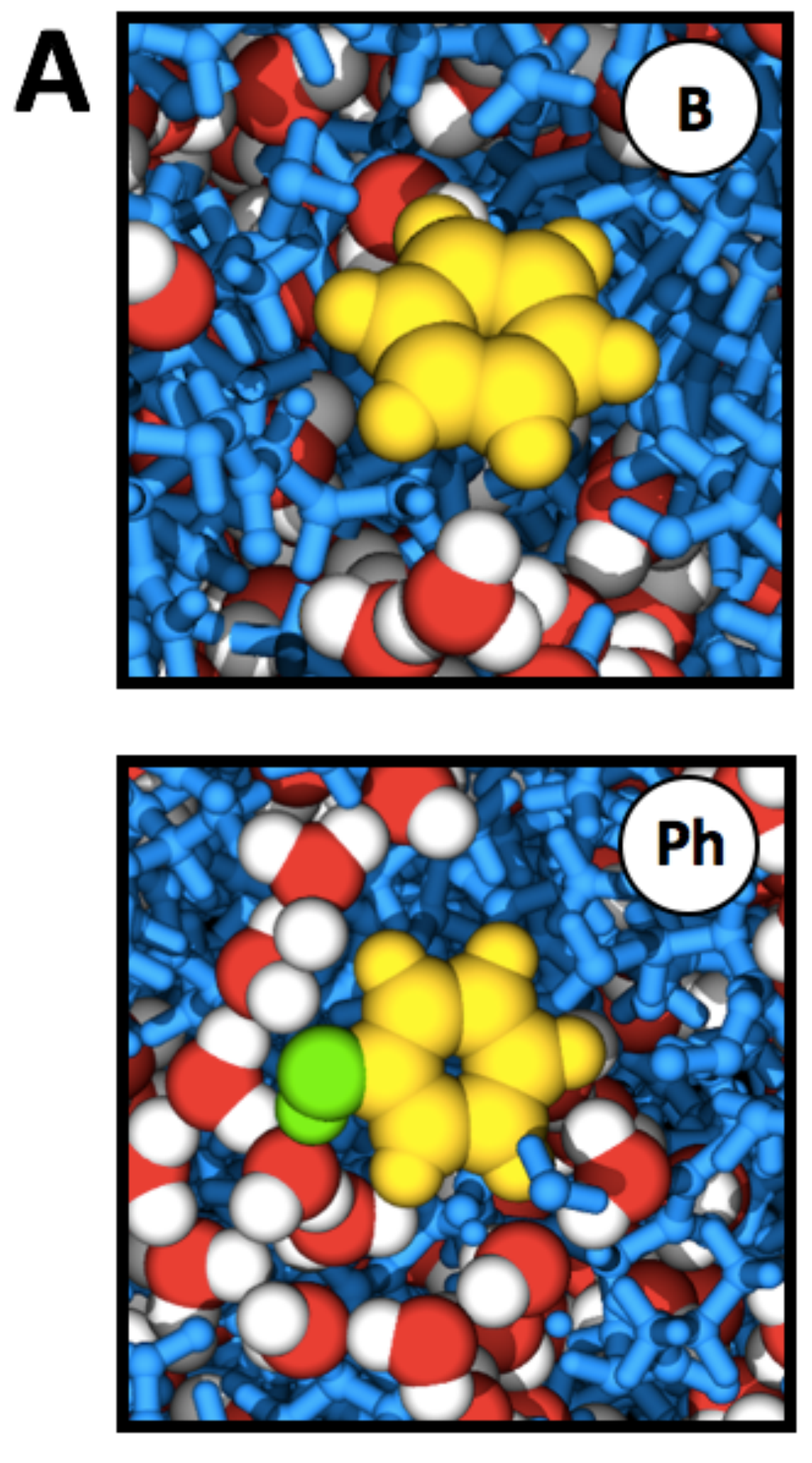}
\end{center}\end{minipage}\hspace{1ex}
\begin{minipage}[b]{0.26\textwidth}\begin{center}
\includegraphics[width=\textwidth]{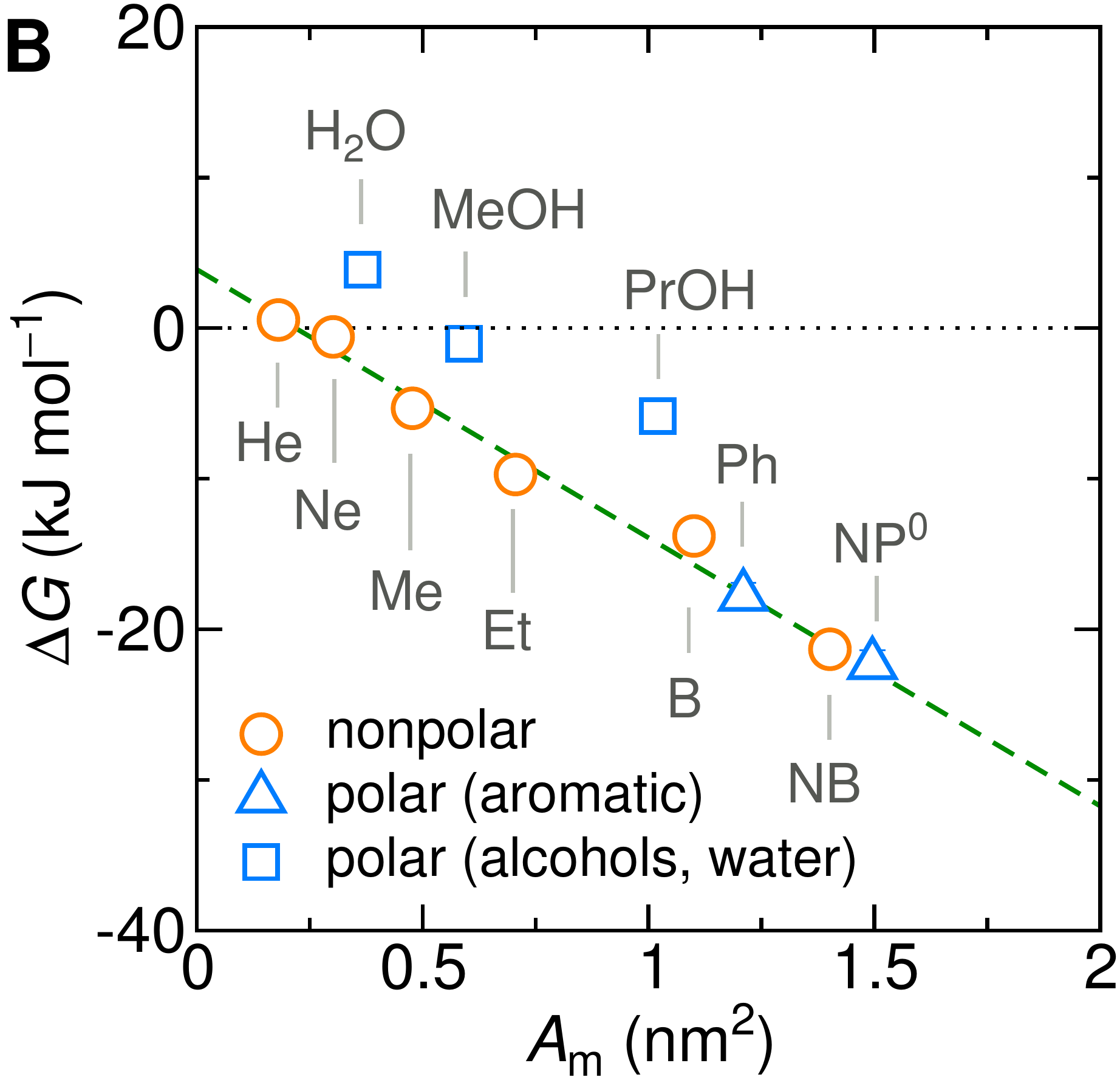}
\end{center}\end{minipage} \hspace{1ex}
\begin{minipage}[b]{0.23\textwidth}\begin{center}
\includegraphics[width=\textwidth]{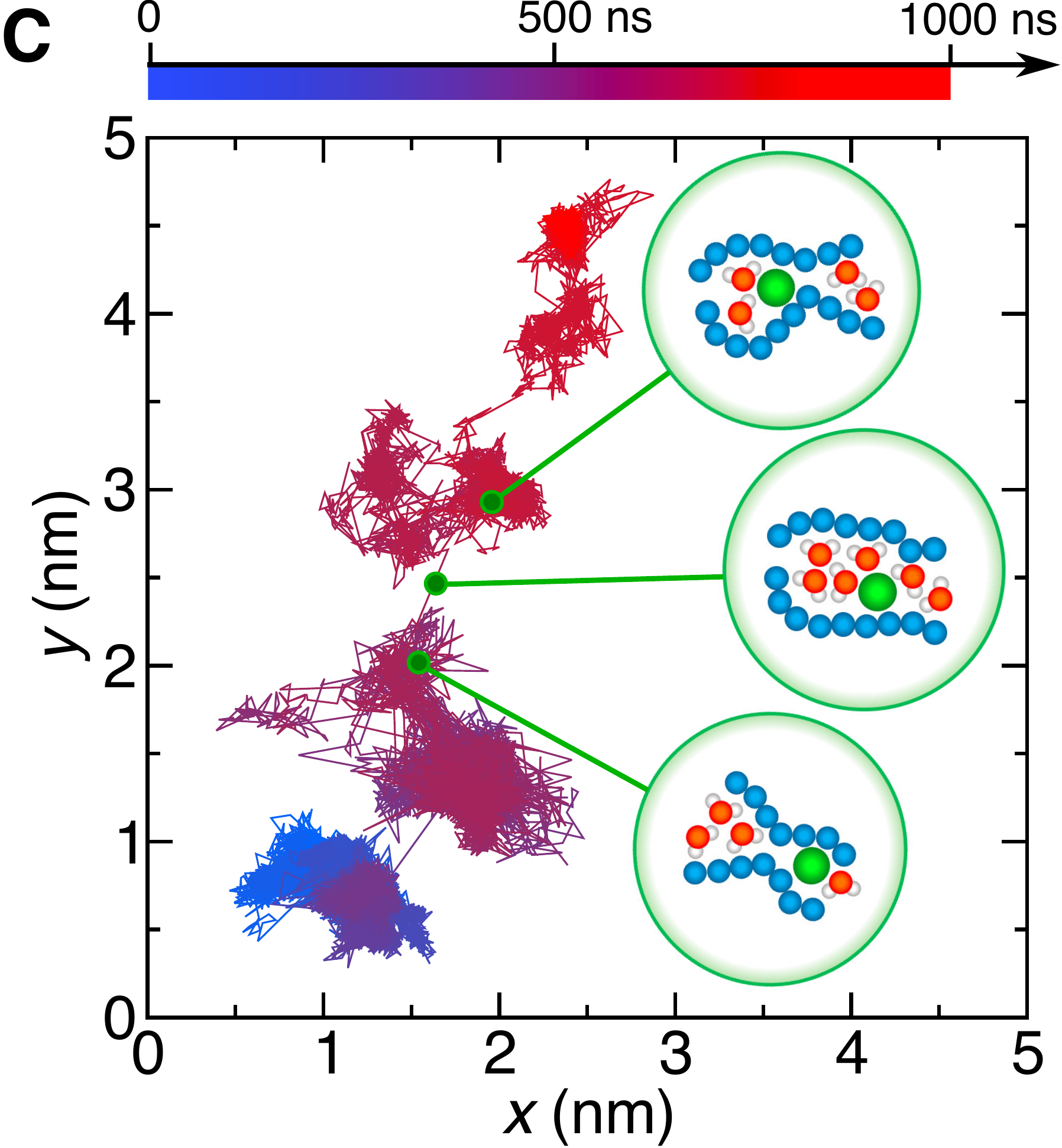}
\end{center}\end{minipage} \hspace{1ex}
\begin{minipage}[b]{0.265\textwidth}\begin{center}
\includegraphics[width=\textwidth]{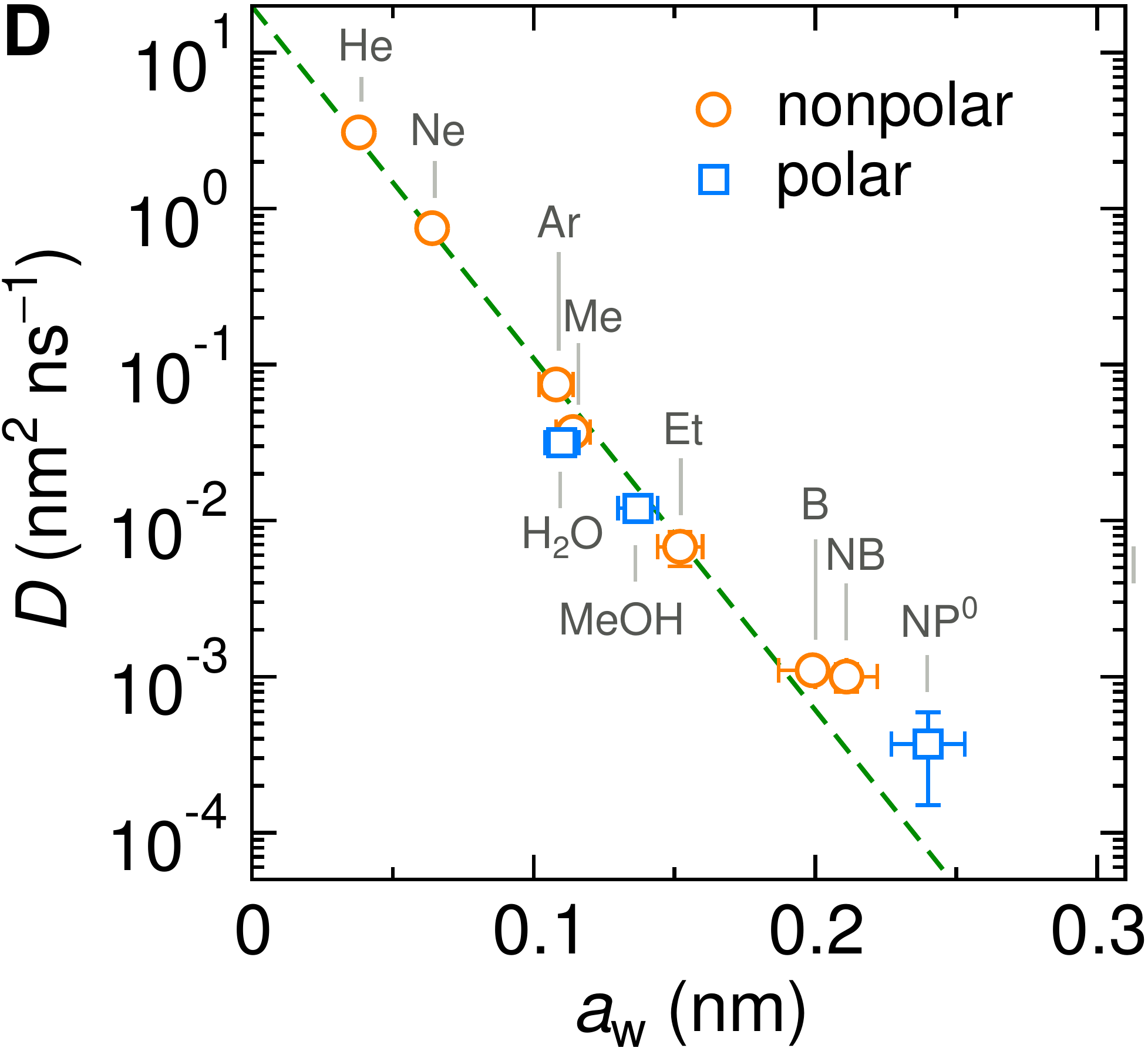}
\end{center}\end{minipage} 
\caption{(A) Snapshots of benzene and phenol molecules solvated in the PNIPAM phase. Hydrophobic parts of the solutes are shown in yellow, the hydroxyl groups in green, PNIPAM polymers in blue, and water in red--white. 
(B) Transfer free energies from water to PNIPAM versus the molecular surface area. The dashed line is a fit of \Eq~(\ref{eq:Ga2}) to the data points of nonpolar solutes.
(C) Microsecond-long trajectory sequence of a NP$^0$ molecule projected on a 2D plane (color coded from blue at $t=0$ to red at $t=1000$~ns). The green bubbles schematically depict the hopping transition with a transient pore opening.
(D) Diffusion coefficients of molecules in the collapsed PNIPAM polymer versus their Stokes radii in water. The dashed line shows a fit of \Eq~(\ref{eq:D_empirical}) to the data points.}
\label{fig:AA-Collapsed}
\end{center}\end{figure*}

Water molecules are very non-uniformly distributed throughout the phase and tend to flock together into irregular clusters of various nanoscopic sizes, which were observed also in simulations of other amorphous polymer structures~\cite{tamai1994molecular, fukuda1998clustering, kucukpinar2003molecular, goudeau2004atomistic,marque2008molecular}.
This water--polymer spatial heterogeneity is a crucial player in the solvation of small molecules,  whereby the nature of the solute (being polar, nonpolar, or ionic~\cite{kanduc2019nano}) is a decisive property. Two representative snapshots in \Fig~\ref{fig:AA-Collapsed}A, showing a benzene (nonpolar) and a  phenol (polar) molecule, demonstrate that nonpolar solutes are preferentially expelled from water clusters and tend to reside in `dryer' regions of the gel, whereas polar molecules tend to partition closer to or inside water clusters. Thus, the ``dual'' character of the gel can favorably accommodate both polar and nonpolar species.

The transfer free energy for a given molecule is obtained as the difference between the solvation free energy in PNIPAM ($G_\trm{g}$) and in water ($G_\trm{w}$),  $\Delta G=\Delta G_\trm{g}-\Delta G_w$, both evaluated via the thermodynamic integration procedure.~\cite{kanduc2019free} Figure~\ref{fig:AA-Collapsed}B shows $\Delta G$ for various solutes plotted versus the molecular surface area $A_\trm m$ of the solutes (defined as the envelope area of the fused union of the atoms~\cite{karelson2000molecular}).
The results follow a clear linear trend for the groups of nonpolar and aromatic solutes as well as alcohols and water. The linearity in the very heterogeneous polymer--water medium
is rather surprising.
The results can be conveniently described in terms of an effective {\sl molecular surface tension} $\gamma_\trm{m}$,~\cite{tanford1979interfacial, ashbaugh2006colloquium}
\begin{equation}
\Delta G=\Delta G_0+\gamma_\trm{m} {A_\trm{m}}.
\label{eq:Ga2}
\end{equation}
$\gamma_\trm{m}$ is strongly related to the difference in the molecule-PNIPAM and molecule--water surface tension. Note that the sign depends on the transfer direction.
The fit of \Eq~(\ref{eq:Ga2}) to the nonpolar solutes (dashed line in \Fig~\ref{fig:AA-Collapsed}B) gives the value $\gamma_\trm{m}=-18$~kJ\,mol$^{-1}$\,nm$^{-2}$.
For the alcohols and water, the transfer free energies are by about 7~kJ/mol above the trend of the nonpolar solutes, owing to a slightly different character of the hydroxyl group than in alkyl chains.~\cite{hansch1991survey}
The molecular size is hence the dominant factor that determines its affinity to the hydrogel. In the CG description (Section 3) the molecular size is therefore reflected in the interaction parameter $\epsilon_\text{ms}$.

Using \Eq~(\ref{eq:partitioning}), we show the partitioning in the collapsed state in \Fig~\ref{fig:AA-K}.
In general, the partitioning of our neutral molecules is larger in the collapsed state. Also, the larger the partitioning in the swollen state, the progressively larger it is in the collapsed state.
With some heuristic arguments, we showed that the partitioning roughly follows the relation $\mathcal{K}_\trm{collapsed}\propto \mathcal{K}_\trm{swollen}^2$.~\cite{kanduc2019free}
This is in line with the universal observation from our CG model (Section 3.1), namely that a collapsed state can have much more extreme effects on partitioning than a swollen state.

Moving on to the diffusion properties of solutes in the collapsed PNIPAM, we first look at the projected trajectory of a NP$^0$ molecule in \Fig~\ref{fig:AA-Collapsed}C.
Its connected blob-like structure suggests that the diffusion advances via the hopping mechanism~\cite{takeuchi1990jump, muller1991diffusion}: 
A penetrating solute resides for longer time in a local cavity and suddenly performs a longer jump into a neighboring cavity through a transient water channel~\cite{kanduc2018diffusion} that forms between the chains (schematically illustrated in the bubbles in \Fig~\ref{fig:AA-Collapsed}C).
We plot the diffusion coefficients versus the size of the solutes $a_\trm{w}$ (defined as the Stokes radius in pure water) in \Fig~\ref{fig:AA-Collapsed}D. 
 As the size of a solute increases by a factor of 7, the diffusion coefficients decreases by dramatic 5 orders of magnitude.
The diffusion coefficients depend on the solute size $a_\trm{w}$ roughly exponentially, 
\begin{equation}
D=D_0\,\rme^{-a_\trm{w}/\lambda}.
\label{eq:D_empirical}
\end{equation}
The fit to the data points yields the decay length $\lambda=0.019$~nm.~\cite{kanduc2018diffusion}

Note that the rate-determining step in the hopping diffusion is the opening of a channel, which is associated with a free energy barrier $\Delta F_\trm{a}$ and can be via Boltzmann probability related to the diffusion coefficient as $D\sim \exp(-\Delta F_\trm{a}/\kB T)$. In conjunction with the empirically obtained diffusion relation [\Eq~(\ref{eq:D_empirical})], this implies
\begin{equation}
\Delta F_\trm{a}(a_\trm w)=\frac{\kB T}{\lambda} a_\trm w.
\label{eq:dF}
\end{equation}
That is, the free energy barrier depends linearly on the particle size, and hence represents a special case of possible scenarios predicted by an assortment of different theories. The majority of theories that are based on activated diffusion predict either square or cubic scaling. 
 A possible linear dependence of the free energy barrier has recently been theoretically envisioned in scaling theories for particle mobility in dense polymer solutions~\cite{cai2011mobility, cai2015hopping} and in dense liquids by using a self-consistent cooperative hopping theory.~\cite{zhang2017correlated}
 As also seen from \Eq~(\ref{eq:dF}) the height of the free energy barrier is related to the decay length $\lambda$ in \Eq~(\ref{eq:D_empirical}). We also showed that in a less hydrated gel, the the diffusion of solutes is lower (i.e., higher $\Delta F_\trm{a}$) and at the same time the diffusion coefficients indeed decay faster with solute size.~\cite{kanduc2018diffusion}

In conclusion, all-atom MD simulations offer insights into the molecular nature of the transport and solvation properties of molecules in hydrogels. These mechanisms are not only important for PNIPAM hydrogels, but most probably play important roles also in other responsive hydrogels, and their understanding is  important for the rational design of novel materials. Notably, we see drastically larger effects for $\cal K$ and $D$ in the collapsed phase than in the swollen states, but apparently also a large anti-correlation between them, like in the coarse-grained simulations in Section~3. {\it The dense, collapsed state is thus more decisive for nanoreactor design and control.} Very recent studies indicate that in particular for charged molecular reactants, the presence of water clusters and resulting substantial interfacial effects within a dense hydrogel may decisively affect their permeability behavior.\cite{kanduc2019nano}  

\section{Concluding remarks}

Stimuli-responsive nanoreactors are of high potential for the design of programmable and selective nano-devices for controlled catalysis and can therefore serve as candidates to create novel synthetic enzymes on the colloidal scale. However, they constitute complex devices with non-equilibrium processes starting at the electronic scale, defining the chemical surface reactions, coupled to those at the polymer network scale with all the intrinsic complexity of polymer--reactant interactions, including the feedback of responsive polymers, up to the device scale where reactants diffuse and react in a suspension of colloids. Here we reviewed the recent theoretical attempts of understanding some  parts of the processes by focussing mostly on the key roles played by the permeability of the polymer shell and the reactant partitioning in order to control activity and selectivity, and how those enter the continuum rate predictions for the nanoreactors. 

As an important general result, we see substantial variations and correlations among $\cal K$ and $D$ in the dense, collapsed polymer phases, in both coarse-grained and atomistically-resolved simulations, which are thus more decisive and tuneable for nanoreactor rate control than for the swollen states. Results for the temperature-induced rate switch observed in reference experiments, like the HCF reduction briefly discussed in Section 2.1, can be thus traced back to, for example, the large exclusion (low partition ratio) and significant slowing down (low diffusion) of reactants in the collapsed state of neutral PNIPAM. However, a quantification of partitioning and diffusion of molecular ions by simulation approaches remains a challenge because of the water heterogeneities in the collapsed states.~\cite{kanduc2019nano}  

A large number of challenges and questions remain, which we try to tackle currently or leave open for future studies. For example, continuum approaches to diffusion- and permeability-influenced rates in confinement are often based on mean-field theories (like presented here), but more elaborate and accurate treatments, like Green's-function approaches,~\cite{benichou}  are yet to be devised. 

The polymer permeability and the knowledge of how reactants partition in the polymer are the keys to program the desired function and response into a nanoreactor. Clearly, the number of experimental and chemical ways to synthesize a responsive hydrogel shell (e.g., with various combinations of copolymerization) is basically infinite. Modeling the features of diverse polymer systems on various scales is therefore out of reach.  Our CG and all-atom studies so far delivered some basic but important insights into the physics of these systems.
However, we are continuing the endeavors towards even more refined notions of the general response features of hydrogels both experimentally and theoretically.  Some of such features are ions, charged reactants,\cite{Herves:2012fp, kanduc2019nano} and even charged (pH-responsive) hydogels,~\cite{Stuart:2010hu} which we ignored in this review, but are of high practical relevance.  Combining all the simulations with continuum-based approaches will help devising models, or at least semi-empirical rules how the hydrogel properties, in particular the permeability of certain molecular species, are connected and can be tuned by stimuli. 

In order to formulate improved rate equations that carry more physical information, also the chemical processes on the nanoparticle surface in the solvent/polymer environment have to be better understood, which we did not touch in this review. For instance, rate-limiting chemical intermediates\cite{Gu:2014cf} could be present. It would be also important to know whether and how strong the (often charged) reactants and products adsorb and diffuse on the nanoparticle surface in the crowded polymer environment. In some cases this may lead to steric hindrance and reaction inhibition on the reactive surface by both reactants and products and highly nonlinear rate behavior.~\cite{toni}  Here, particle-based reaction--diffusion simulations may also help illuminating dynamical transitions and collective effects during the reaction.~\cite{Noe}

Only the fundamental understanding on all scales will enable us to reach the high recognition, selectivity, and feedback behavior in these colloidal devices as found for the nano-sized enzymes.~\cite{enzyme,enzyme2} On the other hand, the large scale and diverse building blocks that constitute the nanoreactors in various architectures establish the opportunity to develop many new design directions within the goal of  programmable, `intelligent' nanoparticle catalysis in the liquid phase. 

\section*{Conflicts of interest}
There are no conflicts to declare

\section*{Acknowledgements}
The authors thank  Richard Chudoba, Karol Palczynski, Sebastian Milster, Arturo Moncho-Jord\'{a}, Stefano Angioletti-Uberti, Daniel Besold, Yan Lu, and Matthias Ballauff for inspiring discussions. This project has received funding from the European Research Council (ERC) under the European Union's Horizon 2020 research and innovation programme (Grant Agreement No. 646659-NANOREACTOR). 
M.K. acknowledges the financial support from the Slovenian Research Agency (research core funding No.\ P1-0055).
W.K.K. acknowledges funding from the Deutsche Forschungsgemeinschaft (DFG) via grant NE 810/11.
The simulations were performed with resources provided by the North-German Supercomputing Alliance (HLRN).

\section*{Author contributions}
All authors contributed equally to this manuscript.

 \setlength{\bibsep}{0pt}

\providecommand*{\mcitethebibliography}{\thebibliography}
\csname @ifundefined\endcsname{endmcitethebibliography}
{\let\endmcitethebibliography\endthebibliography}{}

\bibliographystyle{rsc}

\end{document}